\documentclass[preprint]{revtex4-1}

\usepackage{graphicx}
\usepackage{bm}
\usepackage{scalerel}
\usepackage[utf8]{inputenc}
\usepackage[T1]{fontenc}
\usepackage{newtxtext}
\usepackage{newtxmath}
\usepackage{xcolor}
\usepackage{amsmath}
\usepackage[colorlinks=true, citecolor=blue,
  linkcolor=blue]{hyperref}
\newcommand{\Rey}{Re}
\begin{document}
\setcitestyle{super,open={},close={}}

\title{Dynamics of an internally actuated elastic particle in a plane Poiseuille flow}

\author{Shashikant Verma$^1$, Prateek Anand$^3$ and Navaneeth Kizhakke Marath$^{1,2*}$}
\affiliation{$^1$Department of Mechanical Engineering, Indian Institute of Technology, Ropar 140001, India}
\affiliation{$^2$Centre of Research for Energy Efficiency and Decarbonization (CREED), Indian Institute of Technology Ropar, 140001, India}
\affiliation{$^3$Department of Mechanical Engineering, Indian Institute of Technology Bombay, 400076, India}
\affiliation{\text{$^*$Email address for correspondence: navaneeth@iitrpr.ac.in}}

\date{\today}

\begin{abstract}
We analytically analyse the dynamics of an internally actuated particle, modelled as a compressible elastic sphere embedded with a magnetic bead at its undeformed centre, translating in a plane Poiseuille flow in the Stokes limit. The particle is constrained to translate with a prescribed velocity while remaining at an arbitrary position within the flow by applying an external point force and external point torque at its undeformed centre. The governing equations for the fluid and particle are the Stokes and Navier elasticity equations, respectively. We use the series solutions to the governing equations and the domain perturbation method to capture the deformed shape of the particle, assuming $\alpha \ll 1$. Here, $\alpha$ quantifies the elastic strain induced in the particle due to the viscous stress from the fluid. The external force and external torque are obtained until \textit{O}($\alpha^2$). The particle translating along the channel length experiences an elastic-induced hydrodynamic lift as well as hydrodynamic torque both at \textit{O}($\alpha$) and \textit{O}($\alpha^2$). The leading-order lift depends linearly on the local shear rate and on the combined effects of slip velocity and flow curvature, where the slip velocity is defined as the particle velocity relative to the local ambient flow. The particle reaches a stable equilibrium position away from the centreline, where the net lift vanishes. We show that the direction of deformation‑induced lateral migration of the internally actuated particle is qualitatively distinct from that of drops, capsules, and vesicles in the Stokes limit and from that of rigid spheres undergoing inertial migration.
\end{abstract}

\maketitle

\section{Introduction}
\label{sec:headings}
\begin{figure}
\centerline{\includegraphics[width=0.65\linewidth]{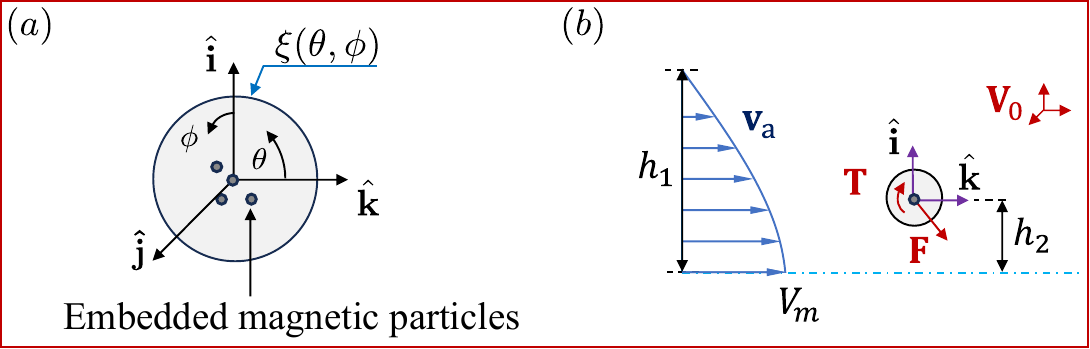}}
  \caption{($a$) Schematic of polymer bead embedded with magnetic nanoparticles \citep{peng2008magnetically}. ($b$) Schematic of an internally actuated elastic sphere translated at $\mathbf{V}_0$ and located at $h_2$ from the centreline in a plane Poiseuille flow ($\mathbf{v}_a$). The half-width of the channel is $h_1$. The origin of the coordinate system is at the undeformed sphere centre, and the sphere is subjected to a point force ($\mathbf{F}$) and a point torque ($\mathbf{T}$). }
\label{fig:problem_schematic}
\end{figure}

An internally actuated elastic particle, such as a smart polymer bead embedded with magnetic nano- or microparticles (MPs), is actuated when an external magnetic field exerts forces on the embedded magnetic particles.
The polymer provides elasticity to the bead, while the embedded MPs provide magnetic responsiveness and offer the advantage of quick and remote actuation \citep{philippova2011magnetic}. Other classes of smart polymer beads also exist, which respond to external stimuli such as light, pH, and temperature \citep{guo2025smart}. Such beads exhibit changes in their properties, structure, or behaviour in response to external stimuli and are used in applications such as tissue engineering \citep{yang2024polysaccharide} and disease diagnosis \citep{tian2024preparation}. The schematic of a polymer bead embedded with MPs is shown in figure \ref{fig:problem_schematic}($a$). The distribution of MPs inside the polymer bead can be either uniform or non-uniform \citep{philippova2011magnetic}. The internally actuated polymer beads are used in many biomedical applications, including in vitro magnetic separation of biological cells, targeted drug delivery and microfluidic-assisted separation of circulating tumour cells \citep{philippova2011magnetic,sung2021magnetic,seyfoori2023microfluidic}. Therefore, it is imperative to understand the dynamics of the beads in a channel flow. 
Given the small size of MPs relative to the bead, the bead experiences localised forces and torques in the presence of an external magnetic field \citep{moerland2019rotating,sung2021magnetic}. The forces and torques can be modelled as point forces and point torques, respectively. Such a model has been used in an analytical study of an elastic sphere (embedded with a single magnetic particle at the undeformed sphere centre) translating parallel to a rigid wall in a quiescent fluid \citep{verma2025dynamics} and in a general unbounded quadratic flow \citep{verma2026dynamics}. 

Understanding the dynamics and morphology of internally actuated particles in microchannels is crucial for both in vivo and in vitro applications. The dynamics of a particle is governed by the properties of the fluid, the particle itself, and the domain boundaries. In the Stokes limit, it is well known that a rigid spherical particle in unidirectional channel flow does not experience hydrodynamic lift due to Stokes reversibility \citep{bretherton1962motion,guazzelli2011physical}. However, lateral migration of a particle can occur due to non-linear effects including inertial effects \citep{segre1961radial,segre1962behaviour,saffman1965lift,saffman1968corrigendum,vasseur1976lateral,hogg1994inertial,nakayama2019three}, particle deformability \citep{leal1980particle}, wall deformability \citep{rallabandi2024fluid}, non-Newtonian fluid behaviour \citep{ho1976migration,chan1977note} and viscous interactions between particles \citep{schonberg1986viscous}. In a Poiseuille flow in the Stokes limit, a deformable particle translating in a Newtonian fluid experiences hydrodynamic lift arising from slip-induced, curvature-induced, and wall-induced mechanisms. In the present analysis, we consider a small deformable particle and restrict our attention to the slip-induced and curvature-induced contributions to the lift. 

The dynamics of deformable particles, such as drops, has been reviewed by \citet{leal1980particle}, capsules and vesicles by \citet{barthes2016motion}; elastic particle suspensions by \citet{villone2019dynamics}; and fluid and elastic solid interactions for near contact regions by \citet{rallabandi2024fluid}. The dynamics of a deformable drop has been analytically analysed in an arbitrary unbounded Stokes flow, with unbounded Poiseuille flow as a particular case \citep{haber1971dynamics}. \citet{haber1971dynamics} have made an error in their calculation and predicted the lateral migration of a neutrally buoyant drop away from the channel centreline. Later, \citet{wohl1974transverse} have corrected the analysis and predicted that the neutrally buoyant drop migrates towards the centreline, in agreement with the experimental results \citep{goldsmith1962flow}. The lateral migration of the drop has been reinvestigated in an unbounded quadratic shear flow and in a linear shear flow bounded by two plane walls \citep{chan1979motion,leal1980particle}. In the quadratic flow, the drop migration can be either toward or away from the centreline, depending on the viscosity ratio of the internal and external fluids of the drop. Whereas in the wall-bounded linear shear flow, the drop always migrates towards the centreline. The migration direction of a non-neutrally buoyant drop can be either towards or away from the centreline, depending on the viscosity ratio, Capillary number (ratio of viscous to surface tension forces) and Bond number (ratio of buoyancy to surface tension forces) \citep{mandal2015effect}.
The centerline stability of particles, including drops, bubbles, and rigid spheres in a cylindrical microchannel, has been investigated experimentally and numerically \citep{cappello2023beads}. A capsule is a drop enclosed by a deformable membrane, whereas a vesicle consists of a drop bounded by a phospholipid bilayer \citep{barthes2016motion}. \citet{kaoui2008lateral} have numerically investigated the lateral migration (towards the centreline) of a two-dimensional vesicle, without viscosity contrast between the inner and outer vesicle fluid, in an unbounded plane Poiseuille flow. \citet{danker2009vesicles} have analytically analysed the motion and deformation of a vesicle and observed that the vesicle migrates towards the centreline in an unbounded plane Poiseuille flow, unlike a drop \citep{leal1980particle}. At a low viscosity ratio (ratio of inner to outer vesicle's fluid viscosity), the vesicle exhibits tank-treading motion, while above a critical viscosity ratio, the vesicle exhibits tumbling or breathing dynamics. At the channel centreline, for the Poiseuille flow of small curvature,  the vesicle exhibits either a bullet-like  (longest axis of the vesicle is oriented in the flow direction) or a parachute-like (longest axis of the vesicle is oriented perpendicular to the flow) shape. While for a large curvature of the flow, the vesicle exhibits a pronounced parachute-like shape \citep{danker2009vesicles}. In an unbounded Poiseuille flow, a capsule also exhibits lateral migration towards the centreline  \citep{helmy1982migration}. \citet{doddi2008lateral} have performed a three-dimensional numerical simulation and observed the lateral migration of a capsule towards the centreline in a plane Poiseuille flow. 

The dynamics of drops, capsules, and vesicles in Poiseuille flow has been extensively studied in the literature, whereas those of an elastic particle and an internally actuated elastic particle remain less explored. Elastic particles serving as biological cell models have been studied for their potential in disease diagnosis \citep{hou2009deformability,villone2019dynamics}. Further, it has been observed that predicting internal stress in the cell is crucial as the stress can alter the expression of certain genes inside the cell and, consequently, determine the cellular properties \citep{zhang2012gene,finney2024impact}. The elasticity of hydrogel particles used in targeted drug applications also influences the drug delivery to a particular target site. In general, harder hydrogel particles are often internalised by immune cells more effectively than softer ones and therefore, hinder the drug transport to the target site \citep{anselmo2017impact}. \citet{tam1973transverse} have analytically analysed the dynamics of a linear elastic compressible sphere suspended in an unbounded simple shear flow in the Stokes limit. The sphere exhibits cross-streamline migration, similar to that of a rigid sphere in an unbounded simple shear flow at finite Reynolds number \citep{saffman1965lift,saffman1968corrigendum}.
The linear elastic solid model has also been used to determine cell dynamics in a real-time deformability cytometry setup \citep{mietke2015extracting}. The leading-order deformed shape of the cell translating along the centreline of a cylindrical channel flow is mapped with that of a linear elastic solid, and the corresponding cell stiffness is quantified. The analytical results of \citet{mietke2015extracting} has been validated using a numerical study on a viscoelastic spherical cell, in the limit of small cell deformation by \citet{mokbel2017numerical}. The cell is surrounded by a thin shell cortex, and the numerical study is also carried out for large deformation of the cell, including either a linear elastic or neo-Hookean hyperelastic model for cell bulk as well as cell cortex. The motion of an incompressible neo-Hookean elastic sphere, initially placed off-centreline in a cylindrical channel, has been studied numerically in both Newtonian and viscoelastic fluids \citep{villone2016numerical}. The sphere migrates towards the centreline in Newtonian fluid, but does not necessarily migrate towards the centreline in the viscoelastic fluid. The deformation of an incompressible neo-Hookean sphere moving along the centerline in Hagen–Poiseuille flow and subjected to an axial body force has been analytically analysed \citep{finney2024impact}. The sphere deforms into a bullet-like shape, anti-bullet shape, or retains sphericity, depending on the strength of the body force and viscous force. The embedding of magnetic particles in an elastic particle results in complex dynamics. For instance, in the absence of an external magnetic field, a compressible magnetic nanogel suspended in a linear shear flow exhibits tumbling and wobbling motion, accompanied by its volume oscillations, as observed numerically \citep{novikau2022behaviour}. The dynamics of an elastic sphere with a centrally embedded magnetic particle has been analysed analytically as the sphere translates parallel to a rigid wall in a quiescent fluid \citep{verma2025dynamics}. The sphere exhibits asymmetric deformation about the translation direction in a plane perpendicular to the wall and tends to migrate away from the wall. The sphere dynamics has also been analysed as it translates in a general unbounded quadratic flow \citep{verma2026dynamics}. To the best of our knowledge, the dynamics of the internally actuated elastic sphere in Poiseuille flow have not been studied analytically.

In the present work, we analysed the dynamics of an internally actuated weakly elastic spherical particle that translates at an arbitrary location in a plane Poiseuille flow, shown in figure \ref{fig:problem_schematic}($b$). We consider that a single magnetic particle is embedded at the centre of the undeformed elastic particle. The response of the magnetic particle to the external magnetic field is modelled using a point force and a point torque, given that the embedded magnetic particle is small in size compared to the overall size of the elastic particle. The particle is modelled as a homogeneous, isotropic, compressible Hookean solid and is constrained to move at an arbitrary location relative to the channel centerline by an external point force and point torque. The force and torque are applied at the centre of its undeformed configuration (sphere). The theoretical framework employed in the present work is similar to that used in the analysis of an internally actuated elastic sphere translating parallel to a rigid wall in a quiescent fluid \citep{verma2025dynamics} and in a general unbounded quadratic flow \citep{verma2026dynamics}. 
The governing equations for the particle and fluid are the Navier elasticity and Stokes equations, respectively. The domain perturbation method and series solutions to the governing equations are adopted to study the particle dynamics. The effects of the channel walls are neglected in the present analysis.

The results of the present work are useful for determining the external force and external torque required to translate an internally actuated particle, located at an arbitrary position relative to the centreline, but away from the channel wall, in plane Poiseuille flow. We find that for the particle translating along the channel length ($z$-direction), the leading-order elastic effects in the point force and point torque comes at \textit{O}($\alpha$). The particle experiences a hydrodynamic lift both at \textit{O}($\alpha$) and \textit{O}($\alpha^2$). We find that there is a location at which net hydrodynamic lift vanishes. Further, the particle, in general, tends to migrate towards (away from) the centreline when it leads (lags) the local ambient flow in the Stokes limit. In contrast, a rigid sphere in Poiseuille flow migrates toward (away from) the centreline in the lagging (leading) case at finite Reynolds number. The lateral migration of the rigid sphere at finite Reynolds number and of the elastic particle in Stokes flow arises as both mechanisms introduce nonlinearity into the system.

The structure of the paper is as follows. In Section \ref{sec:Mathematical formulation}, we present the governing equations, domain perturbation method, modified boundary conditions, solution procedure and expression of velocity, pressure, and displacement fields, and the deformation. The external force and external torque, and the deformed shape of the particle are presented in Section \ref{sec:Analyses of the external force/torque and the deformed shape}. In Section \ref{sec:Comparison of lateral migration with other particles}, we compare the direction of lateral migration of the internally actuated particle with that of a rigid sphere, a spherical drop/capsule/vesicle in Poiseuille flow. In  Section \ref{sec:summary}, we summarise the key observations.

\section{Mathematical formulation}\label{sec:Mathematical formulation}
We consider a weakly elastic spherical particle of undeformed radius $R_0$, translating at a velocity $\mathbf{V}_0$ ($\mathbf{V}_0=V_{0,x}\hat{\mathbf{i}}+V_{0,y}\hat{\mathbf{j}}+V_{0,z}\hat{\mathbf{k}}$) in a plane Poiseuille flow, as shown in figure \ref{fig:problem_schematic}($b$). The maximum flow velocity inside the channel is denoted by $V_m$. The dynamics and morphology of the particle will be symmetric in the upper and lower halves of the channel about the channel centreline ($z$-axis). Therefore, we consider only the upper half of the channel. The half-width of the channel is $h_1$. The particle is located at an arbitrary distance $h_2$ from the channel centreline. Owing to the small particle length scale, we assume that the flow around the particle is governed by viscous effects rather than inertial effects, i.e. Reynolds number, $\Rey = \rho_f V_m R_0^2/(2\mu  h_1) \ll 1$. Here, $\mu$ and $\rho_f$ are the viscosity and density of the fluid, respectively, and $\Rey$ is defined based on the particle length scale. We assume that the particle is located away from the channel walls so that wall effects can be neglected. The fluid is considered Newtonian and incompressible, and its flow is described by the Stokes equations given by
\begin{align}
 \mu \nabla^2  \mathbf{v}- \mathbf{\nabla} p =\mathbf{0},  \label{eq:stokes}
 \end{align}
and the continuity equation given by
\begin{align}
 \mathbf{\nabla}\cdot \mathbf{v}=\mathbf{0}. \label{eq:continuity}
 \end{align}
In (\ref{eq:stokes}) and (\ref{eq:continuity}), the pressure and velocity fields are denoted by $p$ and $\mathbf{v}$, respectively. The particle is modelled as a linear elastic compressible sphere and is described by the Navier elasticity equations, incorporating a point force ($\mathbf{F}$) and a point torque ($\mathbf{T}$), given by
\begin{align}
  \mathbf[\lambda+G]  \mathbf{\nabla}\left( \mathbf{\nabla} \cdot  \mathbf{u}\right)+G  \mathbf{\nabla}^2  \mathbf{u} +\mathbf{F}\delta(\mathbf{x})+\frac{1}{2}\left[\mathbf{\nabla}\delta(\mathbf{x})\times \mathbf{T} \right] =\mathbf{0}.\label{eq:elasticity}
\end{align}
 Here, $\lambda$ and $G$ are the Lam\'e's constants, $\mathbf{u}$ is the displacement field in the particle, and $\delta(\mathbf{x})$ is the Dirac delta function. The fluid stress ($\boldsymbol{\sigma}$) is given by
\begin{align}
 \boldsymbol{\sigma}=-p \mathbf{I}+\mu \left(\nabla \mathbf{v}+[\nabla \mathbf{v}]^T\right) \label{eq:fluid-stress constitutive}
\end{align}
and the particle stress ($\hat{\boldsymbol{\sigma}}$) is given by
\begin{align}
   \hat{\boldsymbol{\sigma}}=\lambda \left(\nabla\cdot\mathbf{u}\right)\mathbf{I}+G\left(\nabla \mathbf{u}+[\nabla \mathbf{u}]^T\right).\label{eq:solid-stress constitutive}
\end{align}
In (\ref{eq:fluid-stress constitutive}) and (\ref{eq:solid-stress constitutive}), $\mathbf{I}$ is the identity matrix. The variables are non-dimensionalised using the particle undeformed radius ($R_0$) and the channel centreline velocity ($V_m$) as the length and velocity scales, respectively. The fluid and particle stresses are non-dimensionalised using the viscous stress ($\mu V_m/R_0$) and shear modulus ($G$), respectively. The point force and point torque are non-dimensionalised using $G R_0^2$ and $G R_0^3$, respectively. The nondimensional variables, indicated using an asterisk symbol,  are obtained as
\begin{align}
    &\mathbf{v}^*=\frac{\mathbf{v}}{V_m};\,\,
    p^*=\frac{p R_0}{\mu V_m}; \,\,\boldsymbol{\sigma}^*=\frac{\boldsymbol{\sigma} R_0}{\mu V_m};\,\,
    \mathbf{u}^*=\frac{\mathbf{u}}{R_{0}};\,\,    \hat{\boldsymbol{\sigma}}^*=\frac{\hat{\boldsymbol{\sigma}}}{G};\,\,\mathbf{F}^*=\frac{\mathbf{F}}{G R_0^2};\,\,\mathbf{T}^*=\frac{\mathbf{T}}{G R_0^3},\nonumber\\& \mathbf{x}^*=\frac{\mathbf{x}}{R_0};\,\,h_1^*=\frac{h_1}{R_0};\,\,h_2^*=\frac{h_2}{R_0};\,\,V_{0,i}^*=\frac{V_{0,i}}{V_m} (\text{where $i=x,y,z$}). \label{eq:nondimensional variables}
\end{align}
Here, $\mathbf{x}$ is the position vector from the undeformed particle centre. The ratio of the bulk modulus ($\lambda$) of the particle to its shear modulus ($G$) is denoted by $\Gamma$; $\Gamma= \lambda/G$. The particle deformability is quantified by the measure of elastic strain induced in the particle by fluid viscous stress, expressed as $\alpha=\mu V_m/(G R_0)$ \citep{finney2024impact,verma2025dynamics}. The parameter $\alpha$ can also be interpreted, either as elastic capillary number, ratio of viscous to elastic stress \citep{villone2019dynamics}, or as the ratio of viscous to elastic force \citep{murata1980deformation,murata1981deformation,nasouri2017elastic}. It can also be interpreted as the ratio of two characteristic timescales ($t_1/t_2$): the particle deformation timescale, $t_1=\mu/G$, and the particle translation timescale, $t_2=R_0/V_0$, where $V_0= |\mathbf{V}_0|$ and $V_0/V_m= O(1)$. In the limit $\alpha \ll1$ or $t_1 \ll t_2$, the particle deforms at a rate much faster than its translational motion. Further, we consider that the flow field near the particle establishes instantaneously as the particle undergoes cross‑stream migration in the Stokes limit. Here onwards, all quantities of interest are expressed in non‑dimensional form, unless specified otherwise. The asterisk symbol used to denote the nondimensional variables in (\ref{eq:nondimensional variables}) is omitted for simplicity.

Considering the nearly spherical shape of the deformed particle (for $\alpha \ll 1$), we express the solutions of (\ref{eq:stokes})-(\ref{eq:elasticity}) in a spherical coordinate system ($\xi$, $\theta$, $\phi$), where $\xi = r/R_0$ is the nondimensional radial coordinate, and $\theta$ and $\phi$ are the polar and azimuthal angles, respectively. The origin of the coordinate system is at the centre of the undeformed particle and translates along with the particle.
The series solutions to (\ref{eq:stokes}) and (\ref{eq:continuity}) are given in Appendix \ref{appsec:Series solution to Stokes continuity equation}, and the solution to (\ref{eq:elasticity}) is given in Appendix \ref{appsec:Series solution to Navier elasticity equation}. The particle rotation is suppressed using the point torque. Therefore, the constants that correspond to pure translation ($c1[0,1]$, $c1[1,1]$, and $c0[1,1]$) and pure rotation with no deformation ($a1[0,1]$, $a1[1,1]$, and $a0[1,1]$) in the series solution given in (\ref{eq:ur series})-(\ref{eq:uphi series}) are set to zero. In the solution outlined in (\ref{eq:ur series})-(\ref{eq:uphi series}), the point force and point torque are represented by $F_{i}$ and $T_{i}$ (where $i=x,y,z$ denotes the Cartesian components), respectively.

We assume the particle attains a steady deformed shape instantaneously during translation, and the surface radius of the deformed particle is given by 
\begin{align}
  \xi=1+f(\theta,\phi). \label{eq:surface}
\end{align}
The surface deformation ($f$) of the particle depends on the polar and azimuthal angles. A form similar to (\ref{eq:surface}) has been employed in the literature to describe the surface of deformable particles, including a deformable drop in a quadratic flow \citep{nadim1991motion} and a weakly elastic sphere in a quiescent fluid \citep{murata1980deformation}. The displacement of a material point from the undeformed surface to its position on the deformed surface is expressed in terms of $f$ as
\begin{align}
    1=\left[1+f(\theta,\phi)-u_r\vert_{\xi=1+f(\theta,\phi)}\right]^2+\left[u_\theta\vert_{\xi=1+f(\theta,\phi)}\right]^2+\left[u_\phi\vert_{\xi=1+f(\theta,\phi)}\right]^2 .\label{eq:deform_disp_relation}
\end{align}
Equation (\ref{eq:deform_disp_relation}) indicates that, for any material point on the deformed surface, the magnitude of the difference between its position on the deformed surface and the displacement it experienced from the undeformed to the deformed surface is equal to one. 

We consider an ambient flow as Poiseuille flow between two rigid infinite parallel plates shown in figure \ref{fig:problem_schematic}($b$). The velocity ($\mathbf{v}_{a}$) and pressure ($p_a$) fields of the flow in terms of nondimensional variables are given by
\begin{align}
    \mathbf{v}_{a}=\left(1-\dot{\gamma}[(h_2/2)+x]-\eta \,x^2\right)\hat{\mathbf{k}}\,,\label{eq:ambient vel field}
\end{align}
and 
\begin{align}
    p_a=P_0-2 \,\xi\,\eta  \cos\theta\,, \label{eq:ambient pressure}
\end{align}
respectively. Here, $\eta=1/h_1^2$ and $\dot{\gamma}=2\,h_2/h_1^2$ are measures of dimensionless flow curvature and local shear rate, respectively. In (\ref{eq:ambient vel field}), $x$ is the distance along the $x$-direction from the origin. The reference pressure at the centre of the undeformed particle is denoted by $P_0$ in (\ref{eq:ambient pressure}). The velocity boundary conditions (no-slip and no-penetration ) at the deformed particle surface is given by
\begin{align}
  \mathbf{v}=\mathbf{0}\,,
  \label{eq:velocity bc}
\end{align}
where $\mathbf{v}=\mathbf{v}_a+\mathbf{v}_{d}-\mathbf{V}_0$. The disturbance velocity field is denoted by $\mathbf{v}_d$. The continuity of traction is the stress boundary condition at the deformed particle surface and is given by
\begin{align}
  \hat{\boldsymbol{\sigma}}\cdot\hat{\mathbf{n}}=\alpha \, \boldsymbol{\sigma}\cdot\hat{\mathbf{n}}\,. \label{eq:stress bc}
\end{align}
Here, $\hat{\mathbf{n}}$ denotes the outward unit normal to the deformed surface, given by $\hat{\mathbf{n}}= (\nabla S/|\nabla S|)$, where $S= \xi-1-f(\theta,\phi)$. The stress boundary condition (\ref{eq:stress bc}) is expressed in terms of the stress components in the spherical coordinate system given by
\begin{align}
  \hat{\sigma}_{rr}-\frac{\hat{\sigma}_{r\theta}}{\xi}\frac{\partial{f}}{\partial{\theta}}-\frac{\hat{\sigma}_{r\phi}}{\xi \sin\theta}\frac{\partial {f}}{\partial{\phi}}=\alpha\left[\sigma_{rr}-\frac{\sigma_{r\theta}}{\xi}\frac{\partial{f}}{\partial{\theta}}-\frac{\sigma_{r\phi}}{\xi \sin\theta}\frac{\partial {f}}{\partial{\phi}}\right]\,,\label{eq:rr stress bc}
\end{align}
\begin{align}
  \hat{\sigma}_{r\theta}-\frac{\hat{\sigma}_{\theta\theta}}{\xi}\frac{\partial{f}}{\partial{\theta}}-\frac{\hat{\sigma}_{\theta\phi}}{\xi \sin\theta}\frac{\partial {f}}{\partial{\phi}}=\alpha\left[\sigma_{r\theta}-\frac{\sigma_{\theta\theta}}{\xi}\frac{\partial{f}}{\partial{\theta}}-\frac{\sigma_{\theta \phi}}{\xi \sin\theta}\frac{\partial {f}}{\partial{\phi}}\right]\label{eq:rtheta stress bc}
\end{align}
and
\begin{align}
  \hat{\sigma}_{r\phi}-\frac{\hat{\sigma}_{\theta\phi}}{\xi}\frac{\partial{f}}{\partial{\theta}}-\frac{\hat{\sigma}_{\phi\phi}}{\xi \sin\theta}\frac{\partial {f}}{\partial{\phi}}=\alpha\left[\sigma_{r\phi}-\frac{\sigma_{\theta\phi}}{\xi}\frac{\partial{f}}{\partial{\theta}}-\frac{\sigma_{\phi \phi}}{\xi \sin\theta}\frac{\partial {f}}{\partial{\phi}}\right]\,.\label{eq:rphi stress bc}
\end{align}
Here, $\hat{\sigma}_{ij}$ and $\sigma_{ij}$ (where $i,j=r,\theta,\phi$) are the components of the solid and fluid stress tensors, respectively, in the spherical coordinate system.
The particle gets deformed as it translates in the Poiseuille flow. The surface deformation ($f$) given in (\ref{eq:surface}) is not known a priori and is determined as a part of the solution. In the weakly deformable limit ($\alpha \ll 1$), the deformed shape of the particle deviates slightly from that of a sphere. Therefore, we employ the domain perturbation method \citep{rangasbook,garyleal} in which the deformed shape is represented as a perturbation about the spherical shape, as discussed in the next section.

\subsection{Domain perturbation method }\label{subsec:domain perturbation method}
The domain perturbation method is applicable to problems in which the flow domain has an irregular geometry, with boundaries that do not correspond to coordinate surfaces of any known analytic coordinate system yet remain close to such surfaces. Since the particle shape deviates slightly from that of a sphere, the boundary conditions are transformed from the deformed surface to the sphere surface ($\xi=1$), yielding modified boundary conditions using the domain perturbation method.  First, we did a Taylor series expansion of the boundary conditions, defined in (\ref{eq:velocity bc})-(\ref{eq:rphi stress bc}) about $\xi=1$, and then substituted the regular asymptotic expansions of the variables ($\mathbf{v}$, $\boldsymbol{\sigma}$, $p$, $\mathbf{u}$, $f$, $\hat{\boldsymbol{\sigma}}$, $\mathbf{F}$, and $\mathbf{T}$) into these expanded boundary conditions. The regular asymptotic expansions of the variables as series in $\alpha$ are given by
\begin{equation}
 \mathbf{v} = \mathbf{v}^{(0)}+\alpha\mathbf{v}^{(1)}+\alpha^2\mathbf{v}^{(2)}+...\, , \label{eq:vel_exp_in_alpha}
\end{equation}
\begin{equation}
 \boldsymbol{\sigma} =\boldsymbol{\sigma}^{(0)}+\alpha\boldsymbol{\sigma}^{(1)}+\alpha^2\boldsymbol{\sigma}^{(2)}+...\, ,\label{eq:fluidstress_exp_in_alpha}
\end{equation}
\begin{equation}
 p =p^{(0)}+\alpha p^{(1)}+\alpha^2 p^{(2)}+...\, ,
\end{equation}
\begin{equation}
 \mathbf{u} =\alpha\mathbf{u}^{(1)}+\alpha^2\mathbf{u}^{(2)}+...\, ,\label{eq:displacement_exp_in_alpha}
\end{equation}
\begin{equation}
f =\alpha f^{(1)}+\alpha^2 f^{(2)}+...\, ,\label{eq:deform_exp_in_alpha}
\end{equation}
\begin{equation}
 \hat{\boldsymbol{\sigma}} =\alpha\hat{\boldsymbol{\sigma}}^{(1)}+\alpha^2\hat{\boldsymbol{\sigma}}^{(2)}+...\, ,\label{eq:solidstress_exp_in_alpha}
\end{equation}
\begin{equation}
 \mathbf{F} =\mathbf{F}^{(0)}+\alpha\mathbf{F}^{(1)}+\alpha^2\mathbf{F}^{(2)}+...
\end{equation}
and
\begin{equation}
 \mathbf{T} =\mathbf{T}^{(0)}+\alpha\mathbf{T}^{(1)}+\alpha^2\mathbf{T}^{(2)}+...\, .\label{eq:torque_exp_in_alpha}
\end{equation} 
Note that the series expansions for the displacement field (\ref{eq:displacement_exp_in_alpha}), surface deformation (\ref{eq:deform_exp_in_alpha}), and solid stress field (\ref{eq:solidstress_exp_in_alpha}) arising from the deformation emerge at \textit{O}($\alpha$) as leading-order terms.
\subsubsection{Modified velocity boundary conditions on the undeformed surface ($\xi=1$)}\label{subsubsec:mod vel bc}
We first expand (\ref{eq:velocity bc})  about the undeformed surface ($\xi=1$) using a Taylor series, substitute the velocity and deformation expansions from (\ref{eq:vel_exp_in_alpha}) and (\ref{eq:deform_exp_in_alpha}), respectively, and then collect terms of equal order in $\alpha$. At \textit{O}(1), the boundary conditions are given by
\begin{align}
  \mathbf{v}^{(0)}&=\mathbf{0} \,\,  \mbox{as $\xi$ =1}\,, \label{eq:vel_v0_bc1} \\
  &\rightarrow \mathbf{v}_a -\mathbf{V}_0 \,\, \mbox{as $\xi \rightarrow \infty$} \,.  
  \label{eq:vel_v0_bc2}
\end{align}
At \textit{O}($\alpha$), it is
\begin{equation} \mathbf{v}^{(1)}+f^{(1)}\frac{\partial{\mathbf{v}^{(0)}}}{\partial{\xi}}=\mathbf{0}   \label{eq:vel_v1_bc}
\end{equation}
and at \textit{O}($\alpha^2$), it is
\begin{equation} \mathbf{v}^{(2)}+f^{(1)}\frac{\partial{\mathbf{v}^{(1)}}}{\partial{\xi}}+f^{(2)}\frac{\partial{\mathbf{v}^{(0)}}}{\partial{\xi}}+\frac{\left[f^{(1)}\right]^2}{2}\frac{\partial^2{\mathbf{v}^{(0)}}}{\partial{\xi}^2}=\mathbf{0} \,.\label{eq:vel_v2_bc}
\end{equation}
The disturbance velocity fields at \textit{O}($\alpha$) [i.e. $\mathbf{v}^{(1)}=\mathbf{v}^{(1)}_d$] and \textit{O}($\alpha^2$) [i.e. $\mathbf{v}^{(2)}=\mathbf{v}^{(2)}_d$] decays as $\xi \rightarrow \infty$. Although the governing equations and the modified velocity boundary conditions at the leading-order are linear, the modified boundary conditions at \textit{O}($\alpha$) and \textit{O}($\alpha^2$) are non-linear due to surface deformation ($f$) being part of the solution, as stated earlier.

\subsubsection{Modified stress boundary conditions on the undeformed surface ($\xi=1$)}\label{subsubsec:mod stress bc}
We first expand (\ref{eq:stress bc})  about the undeformed surface ($\xi=1$) using a Taylor series, substitute the fluid stress, surface deformation, and solid stress from (\ref{eq:fluidstress_exp_in_alpha}), (\ref{eq:deform_exp_in_alpha}), and (\ref{eq:solidstress_exp_in_alpha}), respectively, and then collect terms of equal order in $\alpha$. At \textit{O}($\alpha$), the stress boundary conditions are
\begin{equation}
  \hat{\sigma}^{(1)}_{rr} = \sigma^{(0)}_{rr}\,, \label{eq:Oalpha_stressbc_r}
\end{equation}
\begin{equation}
  \hat{\sigma}^{(1)}_{r\theta} = \sigma^{(0)}_{r\theta} \,,\label{eq:Oalpha_stressbc_theta}
\end{equation}
\begin{equation}
  \hat{\sigma}^{(1)}_{r\phi} = \sigma^{(0)}_{r\phi}\,.\label{eq:Oalpha_stressbc_phi}
\end{equation}
At \textit{O}($\alpha^2$), the boundary conditions are
\begin{equation}
    \hat{\sigma}_{rr}^{(2)}=\sigma_{rr}^{(1)}+f^{(1)} \frac{\partial}{\partial{\xi}}\left[\sigma_{rr}^{(0)}-\hat{\sigma}_{rr}^{(1)}\right]\,,\label{eq:Oalpha^2_stressbc_r}
\end{equation}
\begin{align}
    \hat{\sigma}_{r\theta}^{(2)}=&\sigma_{r\theta}^{(1)}-f^{(1)} \frac{\partial}{\partial{\xi}}\left[\hat{\sigma}_{r\theta}^{(1)}-\sigma_{r\theta}^{(0)}\right]+\frac{\partial{f^{(1)}}}{\partial{\theta}}\left[\hat{\sigma}_{\theta \theta}^{(1)}-\sigma_{\theta \theta}^{(0)}\right] +\frac{1}{\sin\theta}\frac{\partial{f^{(1)}}}{\partial{\phi}}\left[\hat{\sigma}_{\theta \phi}^{(1)}-\sigma_{\theta \phi}^{(0)}\right]\,,\label{eq:Oalpha^2_stressbc_theta}
\end{align}
\begin{align}
    \hat{\sigma}_{r\phi}^{(2)}=&\sigma_{r\phi}^{(1)}-f^{(1)} \frac{\partial}{\partial{\xi}}\left[\hat{\sigma}_{r\phi}^{(1)}-\sigma_{r\phi}^{(0)}\right]+\frac{\partial{f^{(1)}}}{\partial{\theta}}\left[\hat{\sigma}_{\theta \phi}^{(1)}-\sigma_{\theta \phi}^{(0)}\right] + \frac{1}{\sin\theta}\frac{\partial{f^{(1)}}}{\partial{\phi}}\left[\hat{\sigma}_{\phi \phi}^{(1)}-\sigma_{\phi \phi}^{(0)}\right]\,.\label{eq:Oalpha^2_stressbc_phi}
\end{align}
The surface deformation $f^{(1)}$ and $f^{(2)}$ appearing in the modified velocity and stress boundary conditions are obtained by expanding (\ref{eq:deform_disp_relation}) using a Taylor series about $\xi=1$ and subsequently substituting (\ref{eq:displacement_exp_in_alpha}) and (\ref{eq:deform_exp_in_alpha}). At \textit{O}($\alpha$), $f^{(1)}$ is given by
\begin{equation}
    f^{(1)}=\left.u_r^{(1)}\right \vert_{\xi=1}\,. \label{eq:f1_Expression}
\end{equation}
At \textit{O}($\alpha^2$), $f^{(2)}$ is given by
\begin{align}
    f^{(2)}=\left(u_r^{(2)} + f^{(1)} \frac{\partial{u_r^{(1)}}}{\partial{\xi}} -\frac{\left[u_\theta^{(1)}\right]^2 + \left[u_\phi^{(1)}\right]^2}{2}\right)\left. \vphantom{\frac{\left[u_\theta^{(1)}\right]^2}{2}} \right\vert_{\xi=1}\,. \label{eq:f2_Expression}
\end{align}
The series solution to the governing equations, together with the modified boundary conditions, are used to obtain the variables defined in (\ref{eq:vel_exp_in_alpha})-(\ref{eq:torque_exp_in_alpha}).
\subsection{Solution procedure and expressions of field variables}\label{subsec:Solution procedure and expressions of field variables}
In this section, the solution procedure is presented, and expressions for the velocity, pressure, and displacement fields, together with the surface deformation, point force, and point torque, are obtained until \textit{O}($\alpha^2$).
\subsubsection{Fluid velocity and stress fields at leading-order, \textit{O}($1$)}\label{subsubsec:Fluid velocity and stress fields at leading-order}
At the leading-order, the problem reduces to that of a rigid sphere of radius $R_0$, translating at $\mathbf{V}_0$ (where $\mathbf{V}_0=V_{0,x}\hat{\mathbf{i}}+V_{0,y}\hat{\mathbf{j}}+V_{0,z}\hat{\mathbf{k}}$) in the plane Poiseuille flow. The constants ($a[m,n]$, $\tilde{a}[m,n]$, $b[m,n]$, $\tilde{b}[m,n]$, $v[m,n]$, and $\tilde{v}[m,n]$) in the velocity series appearing in (\ref{eq:vr series})-(\ref{eq:vphi series}) of Appendix \ref{appsec:Series solution to Stokes continuity equation} are obtained using the velocity boundary conditions given in (\ref{eq:vel_v0_bc1}) and (\ref{eq:vel_v0_bc2}). The resulting components of the disturbance velocity field ($\mathbf{v}^{(0)}_d$) and the disturbance pressure field ($p^{(0)}_d$) near the particle in the spherical coordinate system are obtained as

\begin{align}
    &v^{(0)}_{d,r}=\frac{A_1}{\xi}+\frac{A_2}{\xi^2}+\frac{A_3}{\xi^3}+\frac{A_4}{\xi^4}+\frac{A_5}{\xi^5},\label{eq:vradial disturb O1_ppflow}\\
    &v^{(0)}_{d,\theta}=\frac{B_1}{\xi}+\frac{B_2}{\xi^2}+\frac{B_3}{\xi^3}+\frac{B_4}{\xi^4}+\frac{B_5}{\xi^5},\\
    &v^{(0)}_{d,\phi}=\frac{C_1}{\xi}+\frac{C_2}{\xi^2}+\frac{C_3}{\xi^3}+\frac{C_4}{\xi^4}+\frac{C_5}{\xi^5}, \label{eq:vphi disturb O1_ppflow}
\end{align}
and
\begin{align}
    &p^{(0)}_{d}=\frac{D_1}{\xi^2}+\frac{D_2}{\xi^3}+\frac{D_3}{\xi^4},
\end{align}
respectively. The $\mathbf{v}^{(0)}_d$ decays as $1/\xi$ as $\xi\rightarrow \infty$. The expressions of $A_i$, $B_i$, $C_i$ (where $i=1,2,...,5$) and $D_i$ (where $i=1,2,3$) are given in Appendix \ref{appsec:expression of constants from A to G}.  The corresponding fluid stress ($\boldsymbol{\sigma}^{(0)}$) is obtained and is used to determine the solid stress ($\hat{\boldsymbol{\sigma}}^{(1)}$) at the next order.
\subsubsection{Solid stress and displacement fields at \textit{O}($\alpha$)}
We obtained the solid stress series from the displacement series given in (\ref{eq:ur series})-(\ref{eq:uphi series}) of Appendix \ref{appsec:Series solution to Navier elasticity equation}. The constants ($a0[m,n]$, $a1[m,n]$, $b0[m,n]$, $b1[m,n]$, $c0[m,n]$, and $c1[m,n]$) in the solid stress series, along with the $\mathbf{F}^{(0)}$ and $\mathbf{T}^{(0)}$ are determined using the stress boundary conditions at \textit{O}($\alpha$) given in (\ref{eq:Oalpha_stressbc_r})-(\ref{eq:Oalpha_stressbc_phi}). The corresponding components of the displacement field ($\mathbf{u}^{(1)}$) are obtained as
\begin{align}
    &u^{(1)}_r=E_1 \xi+E_2 \xi^2+\frac{E_3}{\xi}, \label{eq:ur1_ppflow}\\
    &u^{(1)}_\theta=F_1 \xi+F_2 \xi^2+\frac{F_3}{\xi}+\frac{F_4}{\xi^2},\label{eq:utheta1_ppflow}\\
    &u^{(1)}_\phi=G_1 \xi+G_2 \xi^2+\frac{G_3}{\xi}+\frac{G_4}{\xi^2},\label{eq:uphi1_ppflow}
\end{align}
The expressions of $E_1,E_2,E_3$, $F_1,F_2,F_3,F_4$ and $G_1,G_2,G_3,G_4$ are given in Appendix \ref{appsec:expression of constants from A to G}. Using (\ref{eq:f1_Expression}) and (\ref{eq:ur1_ppflow}), the surface deformation at \textit{O}($\alpha$) is obtained as
\begin{align}
    f^{(1)}=E_1+E_2+E_3. \label{eq:deformation at O alpha_pp flow}
\end{align}
\subsubsection{Fluid velocity and stress fields at \textit{O}($\alpha$)}\label{subsubsec:Fluid velocity and stress fields at Oalpha}
At \textit{O}($\alpha$), the constants ($a[m,n]$, $\tilde{a}[m,n]$, $b[m,n]$, $\tilde{b}[m,n]$, $v[m,n]$, and $\tilde{v}[m,n]$) in the velocity series appearing in (\ref{eq:vr series})-(\ref{eq:vphi series}) of Appendix \ref{appsec:Series solution to Stokes continuity equation} are obtained using the velocity boundary condition given in (\ref{eq:vel_v1_bc}) and the surface deformation ($f^{(1)}$) given in (\ref{eq:deformation at O alpha_pp flow}). The velocity components ($\mathbf{v}^{(1)}_d$) and the pressure field ($p^{(1)}_d$) are obtained as

\begin{align}
    &v^{(1)}_{d,r}=\frac{H_1}{\xi}+\frac{H_2}{\xi^2}+\frac{H_3}{\xi^3}+\frac{H_4}{\xi^4}+\frac{H_5}{\xi^5}+\frac{H_6}{\xi^6}+\frac{H_7}{\xi^7}+\frac{H_8}{\xi^8},\\
    &v^{(1)}_{d,\theta}=\frac{I_1}{\xi}+\frac{I_2}{\xi^2}+\frac{I_3}{\xi^3}+\frac{I_4}{\xi^4}+\frac{I_5}{\xi^5}+\frac{I_6}{\xi^6}+\frac{I_7}{\xi^7}+\frac{I_8}{\xi^8},\\
    &v^{(1)}_{d,\phi}=\frac{J_1}{\xi}+\frac{J_2}{\xi^2}+\frac{J_3}{\xi^3}+\frac{J_4}{\xi^4}+\frac{J_5}{\xi^5}+\frac{J_6}{\xi^6}+\frac{J_7}{\xi^7}+\frac{J_8}{\xi^8},\\
    &p^{(1)}_{d}=\frac{K_1}{\xi^2}+\frac{K_2}{\xi^3}+\frac{K_3}{\xi^4}+\frac{K_4}{\xi^5}+\frac{K_5}{\xi^6}+\frac{K_6}{\xi^7},
\end{align}
The $\mathbf{v}^{(1)}_d$ decays as $1/\xi$ as $\xi\rightarrow \infty$. The expressions of $H_i$, $I_i$, $J_i$ (where $i=1,2,...,8$), and $K_i$ (where $i=1,2,...,6$) are given in supplementary information (SI). The corresponding stress field ($\boldsymbol{\sigma}^{(1)}$) is obtained and is used to determine the solid stress at \textit{O}($\alpha^2$).
\subsubsection{Solid stress and displacement fields at \textit{O}($\alpha^2$)}
At \textit{O}($\alpha^2$), the constants ($a0[m,n]$, $a1[m,n]$, $b0[m,n]$, $b1[m,n]$, $c0[m,n]$, and $c1[m,n]$) in the solid stress series that appear in the corresponding displacement series given in (\ref{eq:ur series})-(\ref{eq:uphi series}) of Appendix \ref{appsec:Series solution to Navier elasticity equation}, and the $\mathbf{F}^{(1)}$ and $\mathbf{T}^{(1)}$ are determined using the stress boundary conditions at \textit{O}($\alpha^2$) given in (\ref{eq:Oalpha^2_stressbc_r})-(\ref{eq:Oalpha^2_stressbc_phi}). The corresponding components of $\mathbf{u}^{(2)}$ are obtained as
\begin{align}
    &u^{(2)}_r=L_1 \xi+L_2 \xi^2+L_3 \xi^3+L_4 \xi^4+L_5 \xi^5+\frac{L_6}{\xi},\label{eq:ur2_ppflow}\\
    &u^{(2)}_\theta=M_1 \xi+M_2 \xi^2+M_3 \xi^3+M_4 \xi^4+M_5 \xi^5+\frac{M_6}{\xi}+\frac{M_7}{\xi^2},\label{eq:utheta2_ppflow}\\
    &u^{(2)}_\phi=N_1 \xi+N_2 \xi^2+N_3 \xi^3+N_4 \xi^4+N_5 \xi^5+\frac{N_6}{\xi}+\frac{N_7}{\xi^2},\label{eq:uphi2_ppflow}
\end{align}
The expressions of $L_i$ (where $i=1,2,...,6$), and $M_i$, $N_i$ (where $i=1,2,...,7$) are given in SI. Using (\ref{eq:f2_Expression}), (\ref{eq:ur1_ppflow})-(\ref{eq:deformation at O alpha_pp flow}), and (\ref{eq:ur2_ppflow}), the surface deformation at \textit{O}($\alpha^2$) is obtained as
\begin{align}
    f^{(2)}=&(E_1 + 2 E_2 - E_3) (E_1 + E_2 + E_3)-\frac{1}{2}( F_1 + F_2 + F_3 + F_4)^2-\frac{1}{2}(G_1 + G_2 + G_3 \nonumber\\&+ G_4)^2 + L_1 + L_2 + L_3 + L_4 + L_5 + L_6. \label{eq:deformation at O^2 alpha_pp flow}
\end{align}
\subsubsection{Fluid velocity and stress fields at \textit{O}($\alpha^2$)}\label{subsubsec:Fluid velocity and stress fields at Oalpha^2}
At \textit{O}($\alpha^2$), the constants ($a[m,n]$, $\tilde{a}[m,n]$, $b[m,n]$, $\tilde{b}[m,n]$, $v[m,n]$, and $\tilde{v}[m,n]$) in the velocity series appearing in (\ref{eq:vr series})-(\ref{eq:vphi series}) of Appendix \ref{appsec:Series solution to Stokes continuity equation} are obtained using the velocity boundary condition given in (\ref{eq:vel_v2_bc}) and are given in SI. The components of the $\mathbf{v}^{(2)}_d$ and the $p^{(2)}_d$ are obtained by substituting the constants in the velocity [(\ref{eq:vr series})-(\ref{eq:vphi series}) of Appendix \ref{appsec:Series solution to Stokes continuity equation}] and pressure series [(\ref{eq:press series}) of Appendix \ref{appsec:Series solution to Stokes continuity equation}], respectively. The $\mathbf{v}^{(2)}_d$ decays as $1/\xi$ as $\xi\rightarrow \infty$. The corresponding fluid stress field ($\boldsymbol{\sigma}^{(2)}$) is obtained and is used to determine the point force and point torque at \textit{O}($\alpha^2$). It is important to note that the point force/torque at a particular order is determined from the stress boundary conditions at the next order. For instance, point force/torque at \textit{O}($1$) is determined from the stress boundary conditions at \textit{O}($\alpha$). So to obtain the point force/torque at \textit{O}($\alpha^2$), one needs to apply the stress boundary conditions at \textit{O}($\alpha^3$). However, in the absence of body force and torque, as in the present work, $\mathbf{F}^{(2)}$ and $\mathbf{T}^{(2)}$ are obtained simply by taking the negatives of the hydrodynamic force and torque derived from the fluid stress at \textit{O}($\alpha^2$), as discussed in detail in Appendix C by \citet{verma2026dynamics}. 

\section{Analyses of the external force, external torque and deformed shape}\label{sec:Analyses of the external force/torque and the deformed shape}
In this section, we discuss the results for the external force, external torque and deformed shape of the particle translating at a distance $h_2$ away from the centreline in the Poiseuille flow.
\subsection{External force acting on the particle}
The external point force needed to translate the particle at a velocity $\mathbf{V}_0$ ($\mathbf{V}_0=V_{0,x}\hat{\mathbf{i}}+V_{0,y}\hat{\mathbf{j}}+V_{0,z}\hat{\mathbf{k}}$) in a plane Poiseuille flow until \textit{O}($\alpha^2$) in dimensional form is obtained as

\begin{align}
    \mathbf{F}=6 \pi \mu V_m R_0(\boldsymbol{\mathcal{F}}^{(0)}+\alpha \boldsymbol{\mathcal{F}}^{(1)}+\alpha^2 \boldsymbol{\mathcal{F}}^{(2)}), \label{eq:total_force_Expression}
\end{align}
\begin{align}
\boldsymbol{\mathcal{F}}^{(0)}=&V_{0,x}\hat{\mathbf{i}}+V_{0,y}\hat{\mathbf{j}}+\left[\frac{\eta}{3} + V_s\right]\hat{\mathbf{k}},\label{eq:O0_force_Expression}\\
\boldsymbol{\mathcal{F}}^{(1)}=&\left[\frac{\dot{\gamma}(\eta+V_s)}{4}-\frac{P_0V_{0,x}}{2+3\Gamma}\right]\hat{\mathbf{i}} -\left[ \frac{P_0 V_{0,y}}{2+3 \Gamma}\right]\hat{\mathbf{j}}\nonumber\\&+\frac{1}{8}\left[\dot{\gamma}V_{0,x}\left(10+\frac{9}{(2+\Gamma)}+\frac{5}{(2+3\Gamma)}\right)-\frac{8P_0(\eta+V_s)}{(2+3\Gamma)}\right]\hat{\mathbf{k}}, \label{eq:O1_force_Expression}\\ 
\boldsymbol{\mathcal{F}}^{(2)}=&\left[S_1+ \frac{P_0(2P_0V_{0,x}-\dot{\gamma}(2\eta+V_s)(2+3\Gamma))}{2(2+3\Gamma)^2} \right]\hat{\mathbf{i}} + \left[S_2+\frac{P_0^2 V_{0,y}}{(2+ 3\Gamma)^2}\right]\hat{\mathbf{j}}\nonumber\\&+ \left[S_3+\frac{P_0((4\eta+2V_s) (2+\Gamma)P_0-\dot{\gamma}V_{0,x}(34+\Gamma(56+15\Gamma)))}{2(2+\Gamma)(2+3\Gamma)^2}\right]\hat{\mathbf{k}}.  \label{eq:O2_force_Expression}
\end{align}
Here, $\eta$ and $\dot{\gamma}$ are defined following (\ref{eq:ambient pressure}), and $V_s=V_{0,z}-[1-(h_2/h_1)^2]$ denote the dimensionless slip velocity. The nondimensional force ($\mathbf{F}/6\pi\mu V_m R_0$) is expressed as sum of forces at \textit{O}($1$), \textit{O}($\alpha$), and \textit{O}($\alpha^2$) represented as $\boldsymbol{\mathcal{F}}^{(0)}$, $\boldsymbol{\mathcal{F}}^{(1)}$ and $\boldsymbol{\mathcal{F}}^{(2)}$, respectively. The expressions of $S_1, S_2$, and $S_3$ in (\ref{eq:O2_force_Expression}) are independent of $P_0$ [defined following (\ref{eq:ambient pressure})] and are given in Appendix \ref{appsec:S1 to S6 expression}. The Cartesian components of $\boldsymbol{\mathcal{F}}$ in the gradient and flow directions are given by $\mathcal{F}_x=\mathcal{F}^{(0)}_x+\alpha\mathcal{F}^{(1)}_x+\alpha^2\mathcal{F}^{(2)}_x$, and $\mathcal{F}_z=\mathcal{F}^{(0)}_z+\alpha\mathcal{F}^{(1)}_z+\alpha^2\mathcal{F}^{(2)}_z$, respectively. The particle can be manipulated to either lead, lag or translate with respect to the local Poiseuille flow velocity depending on the values of $V_{0,x}$, $V_{0,y}$, and $V_{0,z}$. The external force in (\ref{eq:total_force_Expression}) balances the hydrodynamic force acting on the deformed particle to ensure that the particle is force-free in the Stokes limit. The leading-order external force (\ref{eq:O0_force_Expression}) balances the Stokes drag that acts on a rigid sphere of radius $R_0$ translating with velocity $\mathbf{V}_0$ at $h_2$ from centreline in the plane Poiseuille flow. Note that the drag can be obtained using Faxen's law \citep{guazzelli2011physical}. It turns out that the leading-order elastic effect in the external force comes at \textit{O}($\alpha$). The external force at \textit{O}($\alpha$) and \textit{O}($\alpha^2$) scales as $\mu^2 V_m^2/G$ and $\mu^3 V_m^3/(G^2 R_0)$, respectively. 

To analyse the hydrodynamic lift experienced by the internally actuated elastic particle, we consider the particle positioned off‑centreline ($h_2$) and translating only along the $z$-direction, i.e. $V_{0,x}=V_{0,y}=0$. The expressions (\ref{eq:O0_force_Expression})-(\ref{eq:O2_force_Expression}) simplifies to
\begin{align}
    \boldsymbol{\mathcal{F}}^{(0)}_{V_{0,z}}=&\left[\frac{\eta}{3} + V_s\right]\hat{\mathbf{k}}\\
    \boldsymbol{\mathcal{F}}^{(1)}_{V_{0,z}}=&\left[\frac{\dot{\gamma}(\eta+V_s)}{4}\right]\hat{\mathbf{i}}-\left[\frac{P_0(\eta+V_s)}{(2+3\Gamma)}\right]\hat{\mathbf{k}}\label{eq:O1_force_Expression_when_only_V0z}\\
    \boldsymbol{\mathcal{F}}^{(2)}_{V_{0,z}}=&-\left[\frac{P_0\dot{\gamma}(2\eta+V_s)}{4+6\Gamma}\right]\hat{\mathbf{i}}+\left[(S_3)_{V_{0,z}}+\frac{P_0^2(2\eta+V_s)}{(2+3\Gamma)^2}\right]\hat{\mathbf{k}} \label{eq:O2_force_Expression_when_only_V0z}
\end{align}
Here, $\boldsymbol{\mathcal{F}}^{(i)}_{V_{0,z}}=\boldsymbol{\mathcal{F}}^{(i)}$ (for $i=0,1,2$), when $V_{0,x}=V_{0,y}=0$. The expression of $(S_3)_{V_{0,z}}$ appearing in (\ref{eq:O2_force_Expression_when_only_V0z}) is independent of $P_0$ and is given in (\ref{appeq:S_3_V0z}) of Appendix \ref{appsec:S1 to S6 expression}. From (\ref{eq:O1_force_Expression_when_only_V0z}) and (\ref{eq:O2_force_Expression_when_only_V0z}), it can be inferred that the particle experiences an elastic-induced hydrodynamic lift both at \textit{O}($\alpha$) and \textit{O}($\alpha^2$), in addition to a hydrodynamic drag. Note that the lift is not arising from the wall effects but due to interaction with the ambient flow. The lift at \textit{O}($\alpha$) depends on flow curvature ($\eta$), local shear rate ($\dot{\gamma}$) and slip velocity ($V_s$) while the lift at \textit{O}($\alpha^2$) depends on $\eta$, $\dot{\gamma}$, $V_s$, $P_0$ and $\Gamma$. The lift is proportional to $\dot{\gamma}$, and hence vanishes at the channel centerline, despite $\eta$ and $V_s$ being non-zero. The locations at which external force vanishes can be obtained by setting $\mathcal{F}_x=0$. At these locations ($h_{2,e}$'s), the particle does not experience a net hydrodynamic lift. Imposing $\mathcal{F}_x=0$ yields three possible conditions for no lift: $\alpha=0$, or $h_{2,e}=0$, or $h_{2,e}^2=[4 P_0 \alpha + h_1^2 (V_{0,z}-1)(2 P_0 \alpha-2-3\Gamma) -3\Gamma-2]/(2-2 P_0 \alpha+3\Gamma)$. These respectively correspond to the particle being non-deformable, or located on the centreline, or located at an off-centreline equilibrium position. We find that the off-centreline equilibrium position ($h_{2,e}$) approaches the centreline with a decrease in $\alpha$. The hydrodynamic drag at the leading-order depends on $\eta$ and $V_s$, while at \textit{O}($\alpha$) and \textit{O}($\alpha^2$) depends on $\eta$, $V_s$ as well as $P_0$, and $\Gamma$. 

\subsection{External force variation with particle position $h_2$ from centreline}
\begin{figure}
\centerline{\includegraphics[width=0.95\linewidth]{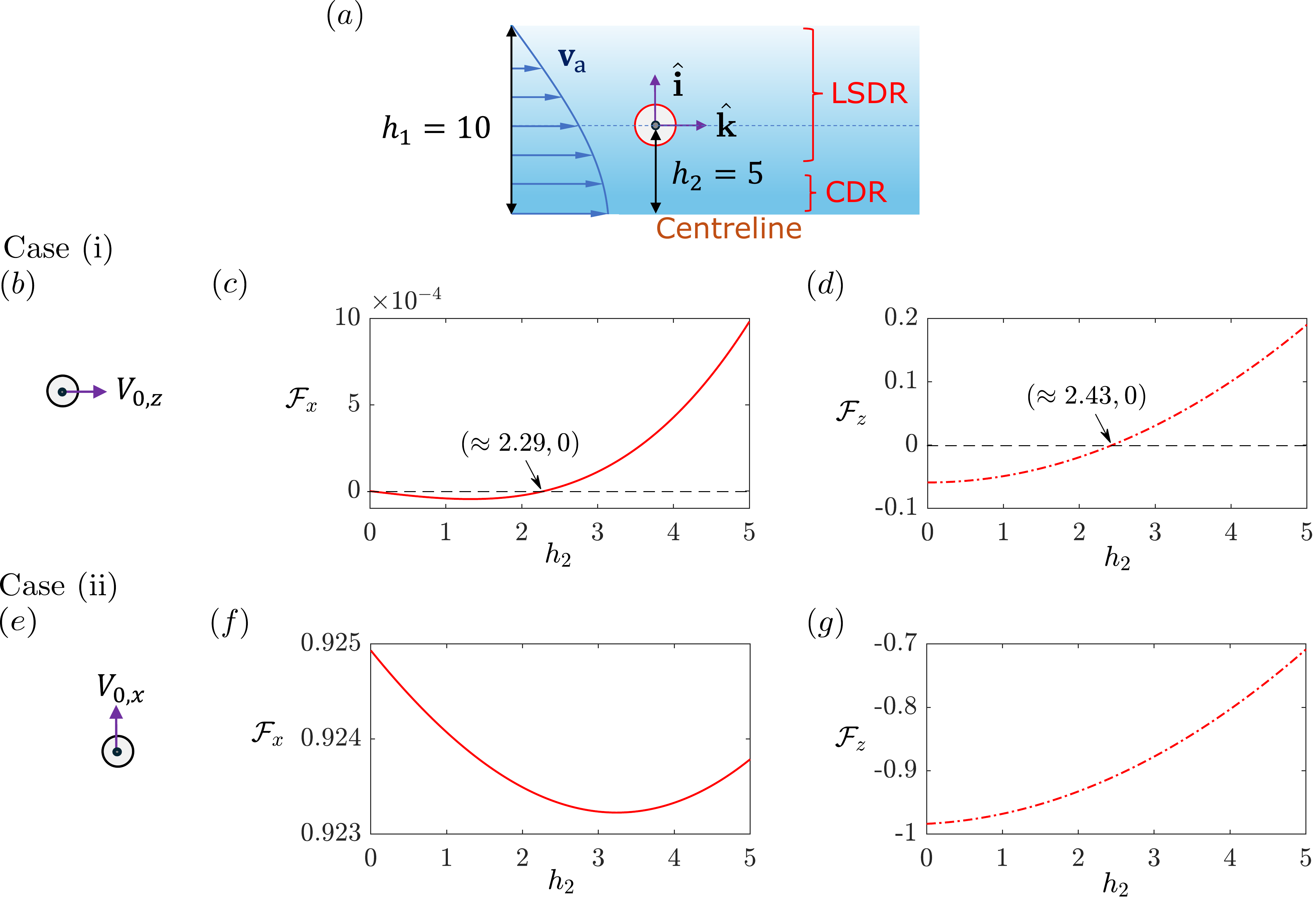}}
  \caption{($a$) Particle translating at a given $h_2$ from the centreline in a channel flow. LSDR: Linear Shear Dominated Region, CDR: Curvature Dominated Region. Case (i): ($b$) Particle translating along the $z$-direction at $V_{0,z}=0.9375$. Variation of the external force ($c$) $\mathcal{F}_x$ and ($d$) $\mathcal{F}_z$ with $h_2$. Case (ii): ($e$) Particle translating along the $x$-direction at $V_{0,x}=0.9375$. Variation of the external force ($f$) $\mathcal{F}_x$ and ($g$) $\mathcal{F}_z$ with $h_2$. Parameters used in the plots; $\Gamma=2$, $P_0=0.1$, $h_1=10$, and $\alpha=0.2$. The Poiseuille flow velocity at $h_2=0.25 h_1$ is $0.9375$. The particle is shown as a sphere in ($a$), ($b$), and ($e$) only for representation.}
\label{fig:force_variation_with_h2}
\end{figure}

As seen in (\ref{eq:O0_force_Expression})-(\ref{eq:O2_force_Expression}), the external force on the particle depends on $h_2$ through $V_s$, $\dot{\gamma}$, $S_1$, $S_2$, and $S_3$. To illustrate the variation of the nondimensional force on the particle with $h_2$  shown in figure \ref{fig:force_variation_with_h2}($a$), we plot $\mathcal{F}_x$ and $\mathcal{F}_z$ with $h_2$ 
for two cases: (i) the particle translates along the $z$-direction (channel length) as shown in figure \ref{fig:force_variation_with_h2}($b$); and (ii) the particle translates along the $x$-direction as shown in figure \ref{fig:force_variation_with_h2}($e$). 
Since we are excluding wall effects, our results are more accurate near the channel centerline. Therefore, we examine the region where $h_2$ varies from $0$ to $0.5\, h_1$ (with $h_1=10$). The results presented in figure \ref{fig:force_variation_with_h2} are obtained using $\alpha=0.2$, $\Gamma=2$, and $P_0=0.1$.

\subsubsection{Case (i): Particle translating in the flow direction}\label{subsubsec:Particle translating in the flow direction}
In case (i), we investigate the forces acting on the particle when it leads or lags the local ambient flow in the region of interest. We constrain the particle to move with $\mathbf{V}_0=0.9375\,\hat{\mathbf{k}}$, which is the ambient flow velocity at $h_2=0.25\times h_1=2.5$. Therefore, the particle leads the local ambient flow for $h_2>2.5$, while it lags the local ambient flow for $0\leq h_2<2.5$. Figure \ref{fig:force_variation_with_h2}(c) shows the variation of the external force in the $x$-direction ($\mathcal{F}_x$) with $h_2$. 
A positive (negative) value of $\mathcal{F}_x$ indicates that the hydrodynamic lift on the particle is directed towards (away from) the centerline. At the centerline ($h_2=0$), $\mathcal{F}_x=0$. As $h_2$ increases, $\mathcal{F}_x$ initially decreases, attaining a negative minimum at $h_2\approx 1.32$, and then increases, vanishing at $h_2\approx 2.29$. With a further increase in $h_2$, $\mathcal{F}_x$ becomes positive and increases monotonically. The non-monotonic variation of $\mathcal{F}_x$ with $h_2$ results from the competition between the slip-induced and curvature-induced lifts acting on the deformable particle, as evident from the terms proportional to $V_s$ and $\eta$, respectively, in (\ref{eq:O1_force_Expression_when_only_V0z}) and (\ref{eq:O2_force_Expression_when_only_V0z}). For a rigid sphere at finite Reynolds number in Poiseuille flow, the slip-induced lift originates from the relative velocity (slip) between the sphere and the surrounding fluid, whereas the curvature-induced lift arises from the curvature of the ambient velocity profile. These two mechanisms lead to lateral forces on the sphere \citep{saffman1965lift,saffman1968corrigendum,hogg1994inertial,matas2004lateral}. In the present problem, although fluid inertia is absent, particle deformability introduces nonlinearity, resulting in slip-induced and curvature-induced lift forces. Their competition leads to the observed non-monotonic dependence of $\mathcal{F}_x$ on $h_2$.

In Poiseuille flow, away from the centerline, the flow field is locally dominated by linear shear, while flow curvature becomes increasingly important near the centerline. Accordingly, the flow may be divided into a linear shear-dominated region (LSDR) and a curvature-dominated region (CDR). A schematic illustrating these two regions is shown in Figure \ref{fig:force_variation_with_h2}($a$). The variation of the lift with $h_2$ can be explained by slip velocity ($V_s$) in LSDR and by curvature effects ($\eta$) in CDR. The magnitude of the slip velocity ($|V_s|$) decreases as we move from channel walls towards $h_2=2.5$, resulting in a decrease in the hydrodynamic lift, hence the external force ($\mathcal{F}_x$). Note that at $h_2=2.5$, the particle translates with the local ambient flow velocity, and the slip velocity is zero, but the lift is non-zero. The non-zero lift results from the flow curvature ($\eta$), as inferred from the $x$-component of the force in (\ref{eq:O1_force_Expression_when_only_V0z}) and (\ref{eq:O2_force_Expression_when_only_V0z}) for $V_s=0$. As $h_2$ decreases from $2.5$, the slip-induced lift should change sign, since the particle now lags the local ambient flow. The net lift vanishes at $h_2 \approx 2.29$, where the slip-induced and curvature-induced lift forces exactly balance each other. As the particle position varies from $h_2 \approx 2.29$ towards the centreline, the magnitude of the slip velocity increases, leading to a corresponding increase in the magnitude of slip-induced lift. As a result, $\mathcal{F}_x$ becomes increasingly negative. This trend continues until $h_2\approx1.32$. Below this $h_2$ value, the strength of the local shear rate decreases, leading to a reduction in the magnitude of slip-induced lift, causing the magnitude of $\mathcal{F}_x$ to decrease. Finally, at the centreline ($h_2=0$), symmetry requires the net hydrodynamic lift to vanish, giving $\mathcal{F}_x=0$.

We find the existence of an off-centreline equilibrium position ($h_{2,e}\approx 2.29$) where the particle experiences no hydrodynamic lift (or $\mathcal{F}_x=0$). For $h_2>2.5$, the particle leads the local ambient flow, and the external force ($\mathcal{F}_x$) is positive, indicating that the hydrodynamic lift on the particle is toward the centreline. However, within the range $0\leq h_2<2.5$, the particle lags the local ambient flow, and the external force ($\mathcal{F}_x$) is negative for $0\leq h_2 \lesssim 2.29$ and positive for $2.29\lesssim h_2<2.5$. Except in the range $2.29\lesssim h_2<2.5$, the direction of the hydrodynamic lift is opposite to that on a rigid sphere (due to inertial effects). For a rigid sphere, leading (lagging) the local ambient flow results in a hydrodynamic lift away from (towards) the channel centreline \citep{vasseur1976lateral,hogg1994inertial}. 

Figure \ref{fig:force_variation_with_h2}(d) shows the variation of the external force in the $z$-direction ($\mathcal{F}_z$) with $h_2$. The force is positive and decreases as $h_2$ decreases from $5$ to approximately $2.5$. This decrease is primarily due to the corresponding reduction in the magnitude of the slip velocity. Physically, the ambient flow tends to advect the particle in the positive $z$-direction. As the magnitude of the slip velocity decreases, the particle velocity approaches that of the surrounding fluid, reducing the external force required to maintain the prescribed particle motion. At $h_2=2.5$, the slip velocity is zero, but $\mathcal{F}_z \neq 0$ due to the flow curvature effects ($\eta$), as inferred from the $z$-component of the force in (\ref{eq:O1_force_Expression_when_only_V0z}) and (\ref{eq:O2_force_Expression_when_only_V0z}) for $V_s=0$. The component $\mathcal{F}_z=0$ at a lower value of $h_2$ i.e. at $h_{2}\approx 2.43$. As the particle position varies from $h_2\approx2.43$ towards the centreline, it lags behind the local ambient flow, and the magnitude of the slip velocity increases. As a result, an external force in the negative $z$-direction is required to maintain the prescribed translational velocity. The magnitude of this force increases as the particle approaches the centreline because the magnitude of the lagging velocity increases.

\subsubsection{Case (ii): Particle translating perpendicular to the flow direction}
In case (ii), the particle translates along the $x$-direction, with $V_{0,z}=V_{0,y}=0$ and $V_{0,x}=0.9375$. The variation of the external force ($\mathcal{F}_x$) with $h_2$ is shown in figure \ref{fig:force_variation_with_h2}($f$). The force remains positive throughout the domain. It decreases with increasing $h_2$, attains a minimum at $h_2\approx 3.24$, and then increases. Since the particle has zero velocity in the flow direction, it lags the local ambient flow in the entire domain. The variation of $\mathcal{F}_x$ with $h_2$ can be explained based on the curvature-induced and slip-induced lateral forces. As explained in section \ref{subsubsec:Particle translating in the flow direction} and figure \ref{fig:force_variation_with_h2}($c$), in the vicinity of the centreline the particle experiences a curvature-induced lateral force away from the centreline. A similar mechanism operates in this case; as $h_2$ increases, the curvature-induced force reduces the magnitude of $\mathcal{F}_x$. The $\mathcal{F}_x$ is minimum at $h_2\approx 3.24$, beyond which slip-induced effects become significant. The magnitude of the slip velocity decreases with increasing $h_2$, and the corresponding slip-induced lateral force diminishes. Therefore $\mathcal{F}_x$  increases to maintain the prescribed particle velocity.

Figure \ref{fig:force_variation_with_h2}($g$) shows the variation of $\mathcal{F}_z$ with $h_2$. As expected, the external force must be applied in the negative $z$-direction to counteract the particle’s tendency to move in the positive $z$-direction due to the ambient flow. As $h_2$ varies from $5$ towards zero at the centreline, the magnitude of the slip velocity increases, leading to a corresponding increase in the absolute value of $\mathcal{F}_z$.

\subsection{Effect of particle deformability on external force}
\begin{figure}
\centerline{\includegraphics[width=\linewidth]{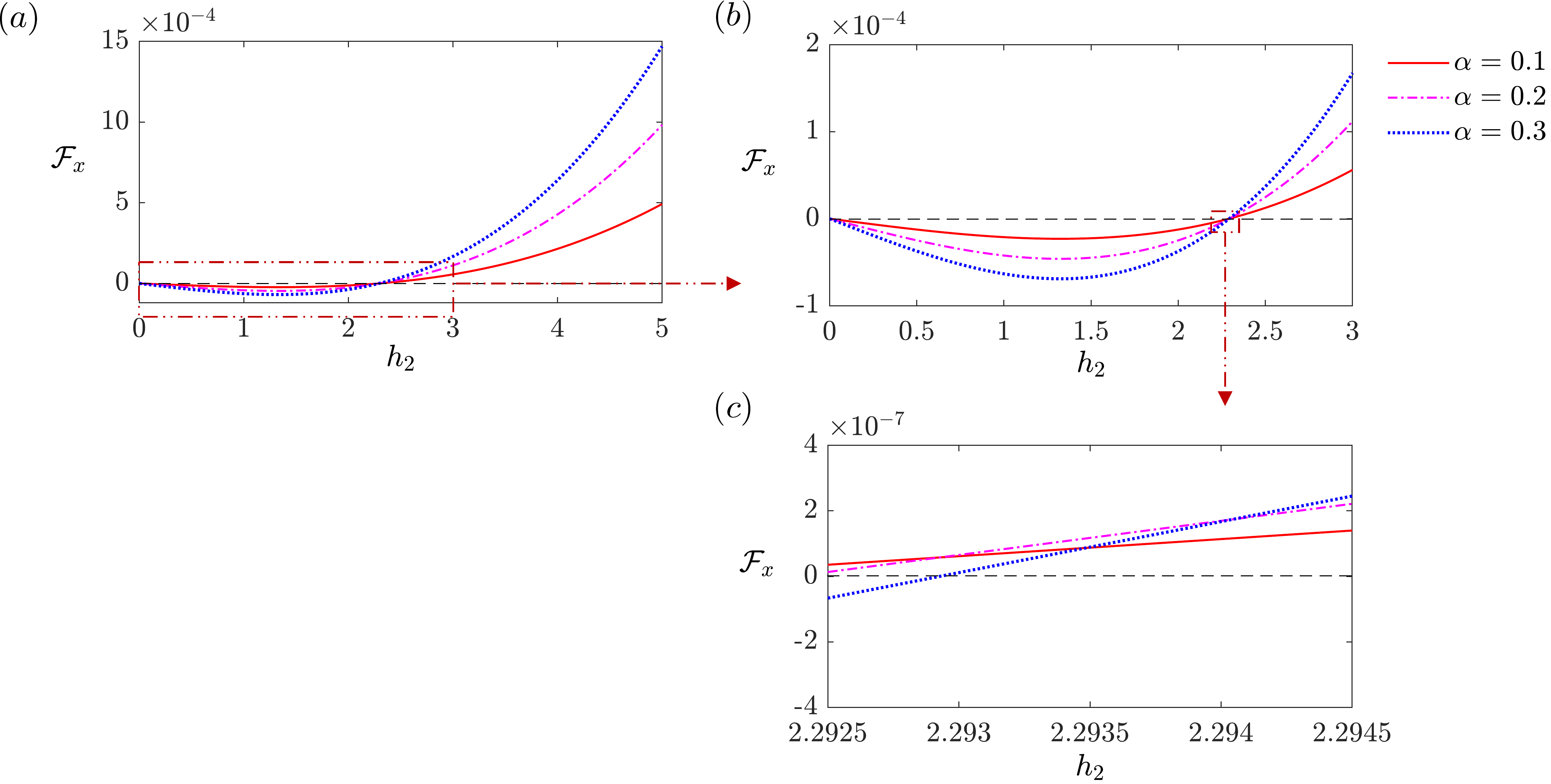}}
  \caption{Variation of external point force ($a$) $\mathcal{F}_x$, with  $h_2$ at different values of $\alpha$, at $V_{0,z}=0.9375$, $V_{0,x}=V_{0,y}=0$, $\Gamma=2$, $P_0=0.1$, and $h_1=10$. The Poiseuille flow velocity at $h_2=0.25 h_1$ is $0.9375$. ($b$) is a subset of ($a$), and ($c$) is a subset of ($b$). Subsets in ($a$) and ($b$) are indicated by a square (dash‑dot‑dot line), not drawn to scale.}
\label{fig:force_x_variation_with_h2_alpha}
\end{figure}

\begin{figure}
\centerline{\includegraphics[width=\linewidth]{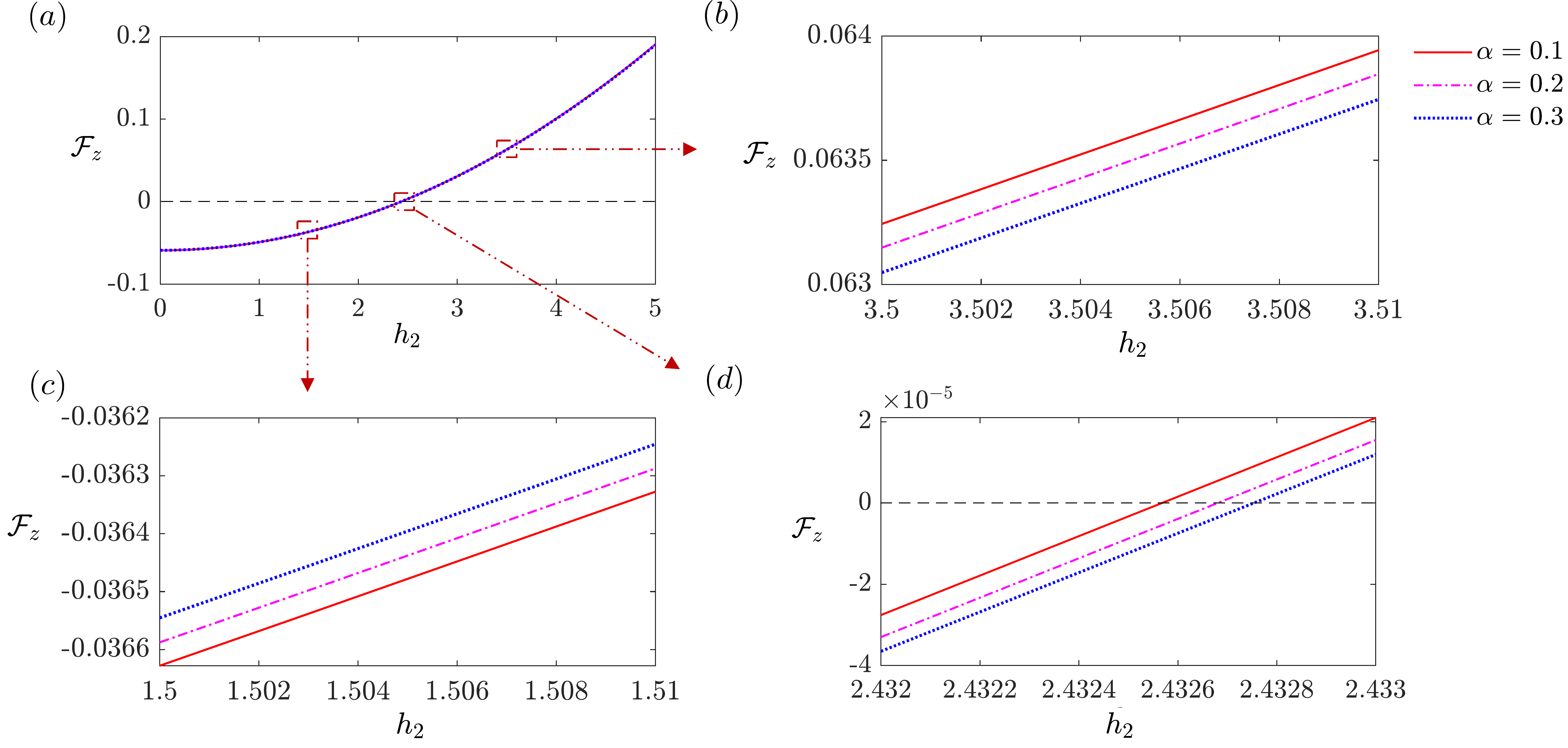}}
  \caption{Variation of external point force ($a$) $\mathcal{F}_z$, with  $h_2$ at different values of $\alpha$, at $V_{0,z}=0.9375$, $V_{0,x}=V_{0,y}=0$, $\Gamma=2$, $P_0=0.1$, and $h_1=10$. The Poiseuille flow velocity at $h_2=0.25 h_1$ is $0.9375$. ($b$), ($c$), and ($d$) are subsets of ($a$). Subsets in ($a$) are indicated by squares (dash‑dot‑dot line), not drawn to scale.}
\label{fig:force_z_variation_with_h2_alpha}
\end{figure}

The particle deformability also influences the external force. We consider the case (i) where $V_{0,z}=0.9375$, and plot the variation of $\mathcal{F}_x$ and $\mathcal{F}_z$ with $h_2$ in figures \ref{fig:force_x_variation_with_h2_alpha} and \ref{fig:force_z_variation_with_h2_alpha}, respectively, for $\Gamma=2$, $P_0=0.1$, $h_1=10$, and $\alpha=0.1,0.2,$ and $0.3$.
Figure \ref{fig:force_x_variation_with_h2_alpha}(b) and (c) show enlarged views of selected regions of figure \ref{fig:force_x_variation_with_h2_alpha}(a).
As seen in the figures, for most values of $h_2$, the magnitude of the external force, $|\mathcal{F}_x|$, increases with particle deformability. This reflects the increase in the magnitude of the hydrodynamic lift with particle deformability. However, we note that this trend is not followed over a narrow range of $h_2$ as seen in figure \ref{fig:force_x_variation_with_h2_alpha}(c). 
Figure \ref{fig:force_z_variation_with_h2_alpha}($a$) shows the variation of $\mathcal{F}_z$ with $h_2$ for different $\alpha$'s. The variation in the absolute value of $\mathcal{F}_z$ is very small for the three $\alpha$'s at a given $h_2$. Therefore, all the three curves of $\mathcal{F}_z$ appear to merge into one another. We note that for $h_2 \gtrsim 2.43$ [figure \ref{fig:force_z_variation_with_h2_alpha}(b)] and $h_2 \lesssim 2.34$ [figure \ref{fig:force_z_variation_with_h2_alpha}(c)], the magnitude of $\mathcal{F}_z$, decreases with increase in particle deformability. However, this trend is not observed in the intermediate range, $2.34 \lesssim h_2 \lesssim 2.43$, as shown in figure \ref{fig:force_z_variation_with_h2_alpha}(d). The corresponding reduction in drag (absolute value) with increasing particle deformability is consistent with the findings of \citet{verma2025dynamics} for a weakly elastic sphere sedimenting parallel to a rigid wall in a quiescent fluid.  

\subsection{Validity of the analysis}\label{subsec:Validity of the analysis}
In our analysis, we have neglected fluid inertia and wall effects. Below, we discuss the limits within which the present analysis remains valid. \citet{hogg1994inertial} derived an expression for the inertial-induced lateral migration velocity of a non-neutrally buoyant rigid sphere translating in a vertical channel flow. The leading-order migration velocity scales as \textit{O}($B R_p^{1/2}$) in the strong shear limit. In this limit, there is a balance between viscous forces and inertial forces arising from the shear flow. Here, $B$ is the buoyancy number, defined as the ratio of the Stokes settling velocity of the sphere to the channel centreline velocity, and in our notation $B=|V_s|$. The particle Reynolds number ($R_p$) is based on the average shear rate of the flow and is defined in our notation as $R_p=\rho V_m R_0^2/(2\mu  h_1)$. Therefore, the present analysis remains valid with fluid inertia neglected, provided that $\alpha^2 \gg B R_p^{1/2}$, so that inertial contributions remain smaller than the deformability contribution.

We have also neglected the wall effects induced by the particle's deformability. As shown by \citet{verma2025dynamics}, the coupling between particle deformability and wall interactions generates a lift force, and the coupling is expected to generate lift in a plane Poiseuille flow as well. We estimate the scaling of the force acting on the particle due to wall interactions. The interactions can be calculated by reflecting the velocity fields $\mathbf{v}^{(0)}_d$, $\mathbf{v}^{(1)}_d$, and $\mathbf{v}^{(2)}_d$ [defined in section \ref{subsec:Solution procedure and expressions of field variables}] from the walls using the method of reflections \citep{kim2013microhydrodynamics,verma2025dynamics}.  
We define the nondimensional distance between the particle centre and the wall as $D$, where $D=h_1-h_2$ for the top wall or $D=h_1+h_2$ for the bottom wall. The velocity field at all orders in $\alpha$ decays as $1/\xi$ as $\xi\rightarrow \infty$, as discussed in sections \ref{subsubsec:Fluid velocity and stress fields at leading-order}, \ref{subsubsec:Fluid velocity and stress fields at Oalpha}, and \ref{subsubsec:Fluid velocity and stress fields at Oalpha^2}. Therefore, the velocity field $\mathbf{v}^{(0)}_d$ after reflecting from the wall scales as $V_m/D$. The reflected velocity field acts as an ambient for the particle, which leads to a force that scales as $\mu V_m R_0/D$, where $D\gg 1$. Similarly, the velocity field at \textit{O}($\alpha^i$) (where $i=1,2$) after reflecting from the wall scales as $\alpha^i V_m/D$, and the corresponding force acting on the particle scales as $(\alpha^i/D)(\mu V_m R_0)$.  In the present work, we obtained the results until \textit{O}($\alpha^2$). The results remain valid even when viscous interactions with the walls are considered, provided $\alpha^2\gg 1/D$.

\subsection{External torque acting on the particle}
The dimensional external torque required to restrict the rotation of the particle located at $h_2$ from the centreline in the plane Poiseuille flow until \textit{O}($\alpha^2$)  is obtained as
\begin{align}
    \mathbf{T}=8 \pi \mu V_m R_0^2(\boldsymbol{\mathcal{T}}^{(0)}+\alpha \boldsymbol{\mathcal{T}}^{(1)}+\alpha^2 \boldsymbol{\mathcal{T}}^{(2)}),\label{eq:total_torque_Expression}
\end{align}
\begin{align}
    \boldsymbol{\mathcal{T}}^{(0)}=&-\dot{\gamma}/2 \,\,\hat{\mathbf{j}},\\
 \boldsymbol{\mathcal{T}}^{(1)}=&\left[\frac{3\eta V_{0,y}(-1+\Gamma)}{(8+12\Gamma)}\right] \hat{\mathbf{i}}+\left[\frac{6P_0\dot{\gamma}(2+\Gamma)-3\eta V_{0,x}(8+\Gamma(17+5\Gamma))}{4(2+\Gamma)(2+3\Gamma)}\right] \hat{\mathbf{j}},\\
 \boldsymbol{\mathcal{T}}^{(2)}=&\left[S_4-\frac{3\eta P_0V_{0,y}(-1+\Gamma)}{(2+3\Gamma)^2}\right] \hat{\mathbf{i}}\nonumber\\&+\left[ S_5  +\frac{3P_0(\dot{\gamma}(2+\Gamma)(-P_0)+\eta V_{0,x}(8+\Gamma(17+5\Gamma)))}{(2+\Gamma)(2+3\Gamma)^2} \right] \hat{\mathbf{j}}+ S_6\,\hat{\mathbf{k}}, \label{eq:external torque at O alpha^2}
\end{align}
The nondimensional torque ($\mathbf{T}/8 \pi \mu V_m R_0^2$) is expressed as sum of torques at \textit{O}($1$), \textit{O}($\alpha$), and \textit{O}($\alpha^2$) represented as $\boldsymbol{\mathcal{T}}^{(0)}$, $\boldsymbol{\mathcal{T}}^{(1)}$ and $\boldsymbol{\mathcal{T}}^{(2)}$, respectively.
The expressions of $S_4$, $S_5$, and $S_6$ in (\ref{eq:external torque at O alpha^2}) are independent of $P_0$ and are given in Appendix \ref{appsec:S1 to S6 expression}. The external torque at \textit{O}($\alpha$) and \textit{O}($\alpha^2$) scales as $\mu^2 V_m^2 R_0/G$ and $\mu^3 V_m^3/G^2$, respectively. The external torque in (\ref{eq:total_torque_Expression}) balances the hydrodynamic torque acting on the deformed particle to ensure that the particle is torque-free in the Stokes limit. The leading-order elastic effect in the torque comes at \textit{O}($\alpha$). 

\subsection{Variation in deformed shape with $h_2$, $V_{0,z}$, $\Gamma$, and $\alpha$}
\begin{figure}
\centerline{\includegraphics[width=0.98\linewidth]{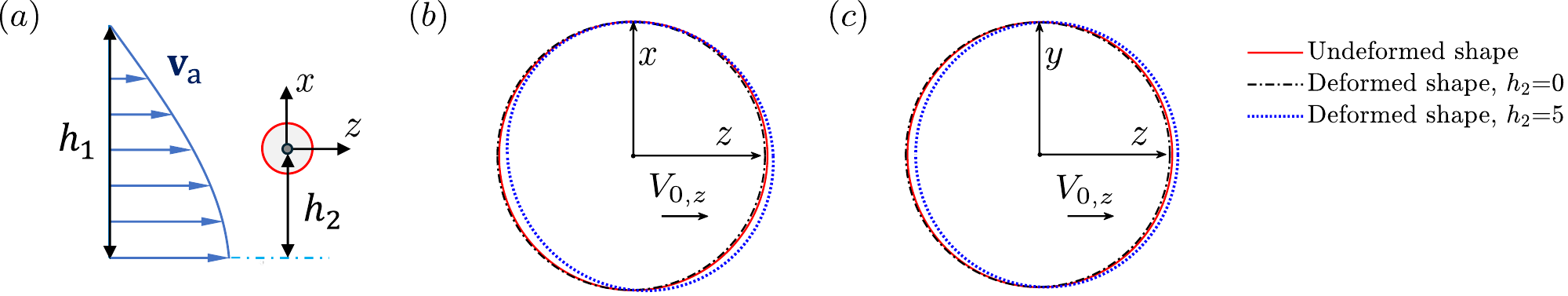}}
  \caption{($a$) Particle translating in the direction of Poiseuille flow ($\mathbf{v}_a$). Variation in the shape of the deformed particle for different values of $h_2$ at $V_{0,z}=0.9375$, $V_{0,x}=V_{0,y}=0$, $\Gamma=2$, $\alpha=0.2$, $P_0=0.1$, and $h_1=10$, plotted on the ($b$) $xz$-plane and ($c$) $yz$-plane. The Poiseuille flow velocity at $h_2=0.25 h_1$ is $0.9375$. The undeformed particle is shown by a solid red line.}
\label{fig:deformed_shape_variation_with_h2}
\end{figure}

\begin{figure}
\centerline{\includegraphics[width=0.9\linewidth]{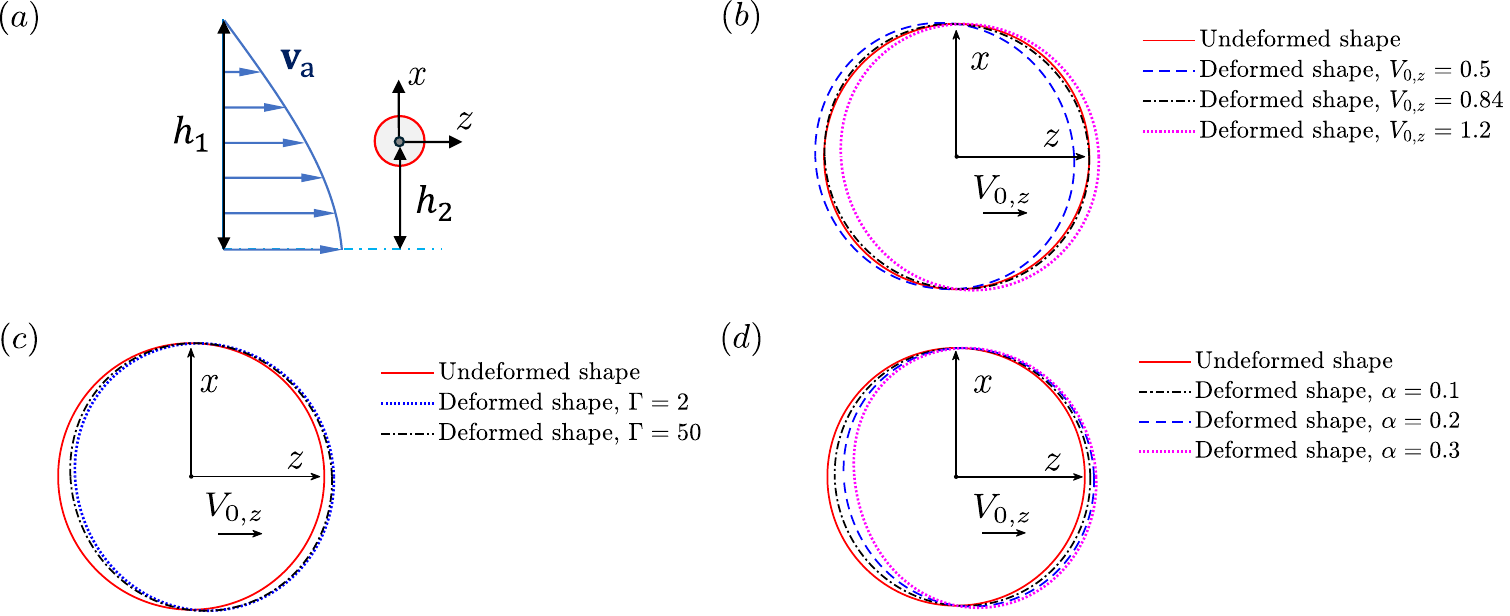}}
  \caption{($a$) Particle translating in the direction of Poiseuille flow ($\mathbf{v}_a$). Variation in the shape of the deformed particle on the $xz$-plane, ($b$) for different values of $V_{0,z}$ at $\Gamma=2$ and $\alpha=0.2$, ($c$) for different values of $\Gamma$ at $V_{0,z}=1.2$ and $\alpha=0.2$, and ($d$) for different values of alpha, at $V_{0,z}=1.2$ and $\Gamma=2$. Parameters: $V_{0,x}=V_{0,y}=0$, $P_0=0.1$, $h_2=4$, and $h_1=10$. The Poiseuille flow velocity at $h_2=0.4 h_1$ is $0.84$. The undeformed particle is shown in a solid red line.}
\label{fig:deformed_shape_variation_with_Vpz_gamma_alpha}
\end{figure}

In this section, we plot the steady-state deformation of the particle for different values of $h_2$, $V_{0,z}$, $\Gamma$, and $\alpha$, while keeping $P_0=0.1$, as it translates along the $z$-direction [figure \ref{fig:deformed_shape_variation_with_h2}($a$)]. The plot highlights deviations of the deformed shape from a perfect sphere. The shapes are plotted for different values of $h_2$ for case (i) [defined in section \ref{subsubsec:Particle translating in the flow direction}] on the $xz$- and $yz$-planes in figure \ref{fig:deformed_shape_variation_with_h2}($b$) and ($c$), respectively.
We consider two values of $h_2$; $h_2=5$, where the particle leads the local ambient flow and $h_2=0$, where the particle lags the local ambient flow. On the $xz$-plane [figure \ref{fig:deformed_shape_variation_with_h2}($b$)], the particle shape is asymmetric about the $z$-axis for all non-zero $h_2$, consistent with the asymmetry of the surrounding flow. The asymmetry in the particle shape becomes pronounced for the particle located away from the centreline. However, on the $yz$-plane [figure \ref{fig:deformed_shape_variation_with_h2}($c$)], the particle shape remains symmetric about the $z$-axis for all $h_2$, consistent with the symmetry of the flow. We plot the deformed shape of the particle translating along the $z$-direction [figure \ref{fig:deformed_shape_variation_with_Vpz_gamma_alpha}($a$)] on the $xz$-plane for different values of $V_{0,z}$, $\Gamma$, and $\alpha$ in figure \ref{fig:deformed_shape_variation_with_Vpz_gamma_alpha}($b$), ($c$), and ($d$), respectively. 
Figure \ref{fig:deformed_shape_variation_with_Vpz_gamma_alpha}($b$) shows the particle shape for $V_{0,z}=0.5$, $0.84$, and $1.2$. The local ambient flow velocity at $h_2=4$ is $0.84$. The deformation is minimal when the particle translates at the local ambient flow velocity, and increases when it leads or lags the flow. 
Figure \ref{fig:deformed_shape_variation_with_Vpz_gamma_alpha}($c$) and ($d$) shows the variation in the particle shape as it translates with $V_{0,z}=1.2$ for different values of $\Gamma$ and $\alpha$, respectively. The deformation increases with decreasing $\Gamma$ and increasing $\alpha$. 

\section{Comparison of lateral migration of internally actuated elastic particles with that of rigid spheres, spherical drops, capsules, and vesicles}\label{sec:Comparison of lateral migration with other particles}
\begin{figure}
\centerline{\includegraphics[width=0.88\linewidth]{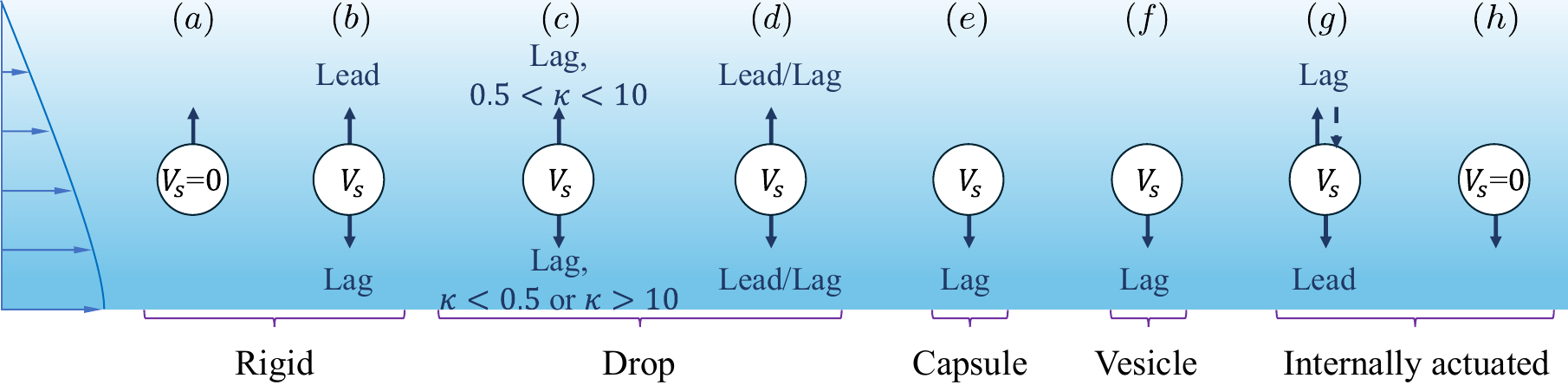}}
  \caption{Lateral migration of ($a, b$) rigid spheres, ($c, d$) spherical drops, ($e$) a capsule, ($f$) a vesicle, and ($g,h$) the internally actutated particles. Particles in ($a$) and ($h$) have zero slip velocity ($V_s=0$), whereas those in ($b$)-($g$) have non-zero slip velocity ($V_s\neq0$). Arrows at the top and bottom of the particles indicate the direction of lateral migration when the particles lead or lag the local ambient flow, as specified by the adjacent text. The dashed arrow in ($g$) indicates the small range of $h_2$ where the lagging internally actuated particle tends to migrate towards the centreline. The shape of a rigid particle and the initial shape of deformable particles is a sphere. Deformable particles are shown as a sphere only for representation.} \label{fig:lateral migration comparison with other particles}
\end{figure}

In this section, we examine the direction of lateral migration of internally actuated particles in comparison with rigid spheres, spherical drops, capsules, and vesicles in Poiseuille flow (figure \ref{fig:lateral migration comparison with other particles}). 
The lateral migration directions of the rigid spheres shown in figure \ref{fig:lateral migration comparison with other particles}($a$) and ($b$) are based on the point-particle approximation, whereas finite-size effects of the particles are considered for deformable particles shown in \ref{fig:lateral migration comparison with other particles}($c$)-($h$). The arrows at the top and bottom of the particles indicate the direction of lateral migration when the particles lead or lag the local ambient flow, as specified by the adjacent text. The particles in figure \ref{fig:lateral migration comparison with other particles}($a$) and ($h$) have zero slip velocity ($V_s=0$), whereas $V_s\neq0$ for the particles in figure \ref{fig:lateral migration comparison with other particles}($b$)-($g$). For the particles in figure \ref{fig:lateral migration comparison with other particles}($b$) and ($d$), the slip velocity is induced by the body force, whereas for those in figure \ref{fig:lateral migration comparison with other particles}($c$)-($h$), it arises due to finite-size effects. For the internally actuated particle in figure \ref{fig:lateral migration comparison with other particles}($g$), the slip velocity is also induced by the external point force. In figure \ref{fig:lateral migration comparison with other particles}($h$), the external point force is applied such that $V_s=0$. Rigid spheres experience lift due to finite fluid inertia, whereas deformable particles experience lift even in the Stokes limit owing to their deformability, as both the inertial and deformability effects introduce nonlinearity into the system. A rigid sphere translating with the local ambient velocity migrates away from the centreline until repelled by the channel walls \citep{matas2004lateral} [figure \ref{fig:lateral migration comparison with other particles}($a$)]. When the sphere has slip velocity, it migrates towards (away from) the centreline if it lags (leads) the local ambient flow \citep{hogg1994inertial} [figure \ref{fig:lateral migration comparison with other particles}($b$)]. The lateral migration of a drop has been analysed in Poiseuille flow \citep{chan1979motion,leal1980particle}. The drop lags the local ambient flow due to its finite size ($V_s\neq0$), and its migration direction is governed by the viscosity ratio ($\kappa$), defined as the ratio of drop viscosity to the suspending fluid viscosity [figure \ref{fig:lateral migration comparison with other particles}($c$)]. When $\kappa$ is between  $0.5$ and $10$, the drop migrates away from the centreline, whereas for other values of $\kappa$, the drop migrates towards the centreline. In the presence of body force induced slip, the direction of lateral migration of a drop does not necessarily depend on whether the drop leads or lags the local ambient flow [figure \ref{fig:lateral migration comparison with other particles}($d$)] \citep{mandal2015effect}. Instead, the lateral migration can occur either towards or away from the centreline, depending on the combined effects of the viscosity ratio, Capillary number (ratio of viscous forces to the surface tension forces), and Bond number (ratio of buoyancy forces to the surface tension forces).
A capsule lags the local ambient flow and migrates towards the centreline \citep{helmy1982migration} [figure \ref{fig:lateral migration comparison with other particles}($e$)]. \citet{doddi2008lateral} have numerically found that a capsule in Poiseuille flow lags the local ambient flow and migrates toward the centreline for viscosity ratios between $0.2$ and $5$, in contrast to the behaviour of the drop \citep{chan1979motion}. \citet{danker2009vesicles} have analytically found that a vesicle lags the local ambient flow and exhibits lateral migration towards the centreline [figure \ref{fig:lateral migration comparison with other particles}($f$)], unlike the drop \citep{chan1979motion,leal1980particle}. In general, the internally actuated elastic particle tends to migrate towards (away from) the centreline as it leads (lags) the local ambient flow [figure \ref{fig:lateral migration comparison with other particles}($g$)]. However, there exist some regions where the lagging particle tends to migrate towards the centreline [dashed arrow in figure \ref{fig:lateral migration comparison with other particles}($g$)], as discussed in section \ref{subsubsec:Particle translating in the flow direction}. Except for these regions, the lateral migration direction of the internally actuated particle with $V_s\neq0$ differs from that of the rigid sphere. Similarly, the internally actuated particle with $V_s=0$ tends to migrate towards the centreline [figure \ref{fig:lateral migration comparison with other particles}($h$)], in contrast to the rigid sphere, which migrates away from the centreline [figure \ref{fig:lateral migration comparison with other particles}($a$)].
 
\section{Summary}\label{sec:summary}
We analyse the dynamics of an internally actuated weakly elastic spherical particle located off-centreline and translated with velocity $\mathbf{V}_0$ ($\mathbf{V}_0=V_{0,x}\hat{\mathbf{i}}+V_{0,y}\hat{\mathbf{j}}+V_{0,z}\hat{\mathbf{k}}$) in a plane Poiseuille flow shown in figure \ref{fig:problem_schematic}($b$). The elastic particle contains a centrally embedded magnetic particle, whose response to an external magnetic field is modelled using a point force and a point torque. The analysis models the response of a polymer bead embedded with a single magnetic particle at its centre. The dynamics of such an internally actuated elastic particle has been analysed as it translates parallel to a rigid wall in a quiescent fluid \citep{verma2025dynamics} and in general unbounded quadratic flow \citep{verma2026dynamics}. We extend the analysis for plane Poiseuille flow while neglecting wall effects. Our results are useful to quantify the external force/torque needed to manipulate the motion of the internally actuated particle located near the centreline in a microchannel.  
We obtain the velocity, pressure, and displacement fields, and the surface deformation, point force, and point torque until \textit{O}($\alpha^2$) in the limit $\alpha \ll 1$. The leading-order elastic effects appears at \textit{O}($\alpha$) in both the external force and external torque given in (\ref{eq:total_force_Expression}) and (\ref{eq:total_torque_Expression}), respectively. At $\textit{O}(\alpha)$ and $\textit{O}(\alpha^2)$, the external force scales as $\mu^2 V_m^2/G$ and $\mu^3 V_m^3/(G^2 R_0)$, while the external torque scales as $\mu^2 V_m^2 R_0/G$ and $\mu^3 V_m^3/G^2$, respectively. The particle located off-centreline and translating along the $z$-direction experiences an elastic-induced hydrodynamic lift in the Stokes limit. The lift until \textit{O}($\alpha^2$) is proportional to the local shear rate ($\dot{\gamma}$) and depends on curvature ($\eta$), slip velocity ($V_s$), and reference pressure ($P_0$), as inferred from the $x$- component of force in (\ref{eq:O1_force_Expression_when_only_V0z}) and (\ref{eq:O2_force_Expression_when_only_V0z}). However, there exists a stable off-centreline equilibrium position ($h_{2,e}$) where the net lift reduces to zero. The equilibrium arises from the balance between the elastic‑induced lift generated by the slip velocity and the curvature effects of the Poiseuille. Note that such an equilibrium position is absent for drops, capsules, and vesicles, as well as for rigid spheres (where migration is driven by inertial effects) in the Poiseuille flow.
We analyse the variation of external force ($\mathcal{F}_x$ and $\mathcal{F}_z$) with $h_2$ for two cases: (i) particle translating along the flow direction, and (ii) particle translating perpendicular to the flow, shown in figure \ref{fig:force_variation_with_h2}. We find that in the first case, for most of the $h_2$, the particle tends to migrate towards the channel centreline when it leads the local ambient flow, whereas away from the centreline when it lags the local ambient flow, as discussed in section \ref{subsubsec:Particle translating in the flow direction}. In the Stokes limit, the particle experiences a hydrodynamic lift arising from the effects of slip velocity and the curvature of the Poiseuille flow, analogous to the inertial lift experienced by a rigid sphere at finite Reynolds number. 
The variation of $\mathcal{F}_x$ and $\mathcal{F}_z$ with $h_2$ is shown in figure \ref{fig:force_variation_with_h2} ($c,d$), and ($f,g$) for the first and second cases, respectively. Figure \ref{fig:force_x_variation_with_h2_alpha} shows that, for most of the $h_2$, the highly deformable particle translating along the $z$-direction, experiences a large hydrodynamic lift (absolute value), hence a larger $|\mathcal{F}_x|$ is needed to constrain the particle motion. However, within a narrow region, $\mathcal{F}_x$ exhibits a non-monotonic variation with respect to $\alpha$ [figure \ref{fig:force_x_variation_with_h2_alpha}($c$)]. Figure \ref{fig:force_z_variation_with_h2_alpha} shows the variation of $\mathcal{F}_z$ with $h_2$ for different $\alpha$'s. The $|\mathcal{F}_z|$ decreases with an increase in deformability for a certain range of $h_2$, which indicates that the highly deformable particle experiences less hydrodynamic drag (absolute value). The present analysis remains valid even while considering the inertial effects provided $\alpha^2 \gg B R_p^{1/2}$ and the viscous interactions with the wall provided $\alpha^2\gg 1/D$, $\alpha^2\gg 1/(L-D)$, as discussed in section \ref{subsec:Validity of the analysis}. We plot the deviation of the deformed particle shape from that of a perfect sphere as the particle translates along the flow direction. The variation in the deformed shape with $h_2$ is shown in figure \ref{fig:deformed_shape_variation_with_h2}, while that with $V_{0,z}$, $\Gamma$, and $\alpha$ is shown in figure \ref{fig:deformed_shape_variation_with_Vpz_gamma_alpha}. The shape is asymmetric about the $z$-axis on the $xz$-plane for non-zero value of $h_2$ [figure \ref{fig:deformed_shape_variation_with_h2}($b$)] while symmetric on the $yz$-plane for all values of $h_2$ [figure \ref{fig:deformed_shape_variation_with_h2}($c$)]. The particle exhibits large deformation when it either leads or lags the local ambient flow compared to when it translates with the local ambient flow velocity [figure \ref{fig:deformed_shape_variation_with_Vpz_gamma_alpha}($b$)]. Further, the particle deformation increases with decrease in $\Gamma$ [figure \ref{fig:deformed_shape_variation_with_Vpz_gamma_alpha}($c$)] and increase in $\alpha$ [figure \ref{fig:deformed_shape_variation_with_Vpz_gamma_alpha}($d$)]. We contrast the lateral migration of the internally actuated particle with that of a rigid sphere, spherical drop/capsule/vesicle in section \ref{sec:Comparison of lateral migration with other particles}. 
Except for certain regions, the deformation-induced lateral migration direction of the internally actuated elastic particle differs from that of the inertial migration of a rigid sphere in the Poiseuille flow. The overall dynamics and steady-state morphology of the internally actuated elastic particle depend on the compressible nature of the particle and the force distribution acting on it. The theoretical framework presented in this work can be extended to account explicitly for the channel wall effects using the method of reflections, although the scaling analysis in section \ref{subsec:Validity of the analysis} identify the parameter regimes in which the present results remain valid for wall-bounded Poiseuille flow. The method has been used to capture the wall effects for the particle translating parallel to a rigid wall in a quiescent fluid \citep{verma2025dynamics}.\\
\textbf{Supplementary data.} The expressions of $H_i$, $I_i$, and $J_i$ (where $i=1,2,...,8$), $K_i$ and $L_i$ (where $i=1,2,...,6$), and $M_i$ and $N_i$ (where $i=1,2,...,7$); and the constants ($a[m,n]$, $\tilde{a}[m,n]$, $b[m,n]$, $\tilde{b}[m,n]$, $v[m,n]$, and $\tilde{v}[m,n]$) required to determine the disturbance velocity and pressure fields at \textit{O}($\alpha^2$) are provided in the supplementary material (SI, Wolfram Mathematica sheet). \\
\textbf{Funding.} This work was supported by the Department of Science and Technology, India (SERB-MATRICS Scheme [MTR/2023/001354] and Core Research Grant [CRG/2023/004878]), and by the Prime Minister's Research Fellowship (PMRF), Ministry of Education, India. \\
\textbf{Declaration of interests.} The authors report no conflict of interest.\\
\textbf{Author ORCIDs.}{ Shashikant Verma, http://orcid.org/0009-0006-5566-8552; Prateek Anand, https://orcid.org/0000-0001-6922-0894; Navaneeth Kizhakke Marath,  http://orcid.org/0000-0001-8769-2633.}
\appendix
\section{Series solutions to the Stokes and continuity equations}\label{appsec:Series solution to Stokes continuity equation}
 The radial ($v_r$), tangential ($v_\theta$), and azimuthal ($v_\phi$) components of the velocity field are given by
 \begin{align}
v_r=&\sum_{n=1}^{\infty} \sum_{m=0}^{n} \left[\left (\frac{\xi^{-n}(n+1)}{2(2n-1)}P_n^m[a[m,n] \cos m\phi + \tilde{a}[m,n] \sin m\phi]\right)\right.\nonumber\\& \left.- 
\vphantom{\frac{(n+1)}{2(2n-1)}}\left(\xi^{-n-2}(n+1)P_n^m[b[m,n] \cos m\phi +\tilde{b}[m,n] \sin m\phi]\right)\right]\,,\label{eq:vr series}
\end{align}
\begin{align}
v_{\theta}=&\sum_{n=1}^{\infty} \sum_{m=0}^{n}\left[\left(\frac{\xi^{-n}(n-2)}{2n(2n-1)(2n+1) \sin \theta}[(n+1)(n+m)P_{n-1}^m\right.\right.\nonumber\\&\left.\left.-n(n-m+1)P_{n+1}^m]\vphantom{\frac{\xi^{-n}(n-2)}{2n(2n-1)(2n+1) \sin \theta}}\right)\left[\vphantom{\frac{\xi^{-n}(n-2)}{2n(2n-1)(2n+1) \sin \theta}}a[m,n] \cos m\phi + \tilde{a}[m,n] \sin m\phi\right]\right.\nonumber\\&\left.-\left (\frac{\xi^{-n-2}}{(2n+1) \sin \theta} [(n+1)(n+m)P_{n-1}^m-n(n-m+1)P_{n+1}^m]\right)\right.\nonumber\\&\left. \times \left[\vphantom{\frac{\xi^{-n-2}}{(2n+1) \sin \theta}}b[m,n] \cos m\phi + \tilde{b}[m,n] \sin m\phi\right]\right.\nonumber\\&\left.+
\left (\frac{\xi^{-n-1} m}{\sin \theta}P_n^m[-v[m,n] \sin m\phi +\tilde{v}[m,n] \cos m\phi]\vphantom{\frac{\xi^{-n-1} m}{\sin \theta}}\right)\right]\label{eq:vtheta series}
\end{align}
and
\begin{align}
v_{\phi}=&\sum_{n=1}^{\infty} \sum_{m=0}^{n}\left[\left(-\frac{\xi^{-n}(n-2)m}{2n(2n-1)\sin \theta} P_n^m [-a[m,n] \sin m\phi + \tilde{a}[m,n] \cos m\phi]\vphantom{\frac{\xi^{-n}(n-2)m}{2n(2n-1)\sin \theta}}\right)\right.\nonumber\\&\left.+
\left (\frac{\xi^{-n-2} m}{\sin \theta} P_n^m [-b[m,n] \sin m\phi +\tilde{b}[m,n] \cos m\phi]\right)\right.\nonumber\\&\left.+
\left (\frac{\xi^{-n-1}}{(2n+1)\sin \theta}[(n+1)(n+m)P_{n-1}^m -n(n-m+1)P_{n+1}^m]\right)\right.\nonumber\\&\left.\times (v[m,n] \cos m\phi +\tilde{v}[m,n] \sin m\phi)\vphantom{\frac{\xi^{-n}(n-2)m}{2n(2n-1)\sin \theta}}\right]\,,\label{eq:vphi series}
\end{align}
respectively. The pressure field is given by
\begin{align}
p&= \sum_{n=1}^{\infty} \sum_{m=0}^{n}\left[\xi^{-n-1}P_n^m(a[m,n] \cos m\phi + \tilde{a}[m,n] \sin m\phi)\right] \,. \label{eq:press series}
\end{align} 
Above, $P^m_n=P^m_n(\cos \theta)$ is the associated Legendre polynomial of order $m$ and degree $n$. The constants, $a[m,n]$, $\tilde{a}[m,n]$, $b[m,n]$, $\tilde{b}[m,n]$, $v[m,n]$, and $\tilde{v}[m,n]$ are determined at different orders in $\alpha$ until \textit{O}($\alpha^2$) from the corresponding modified velocity boundary conditions, as discussed in section \ref{subsubsec:mod vel bc}.

\section{Series solutions to the Navier elasticity equations}\label{appsec:Series solution to Navier elasticity equation}
The radial ($u_r$), tangential ($u_{\theta}$), and azimuthal ($u_\phi$) components of the displacement field are given by
\begin{align}
u_r=&\frac{1}{4 \pi \xi}\left[\vphantom{\frac{1}{4 \pi \xi}}F_{z} \cos \theta + F_{x} \cos \phi \sin \theta + F_{y} \sin \theta \sin \phi\right]\nonumber \\&+\sum_{n=0}^{\infty} \sum_{m=0}^{n}\left[\xi^{n+1}P_n^m(b1[m,n] \cos m\phi +b0[m,n] \sin m\phi)\right] \nonumber \\&+\sum_{n=1}^{\infty} \sum_{m=0}^{n}\left[\xi^{n-1}P_n^m(c1[m,n] \cos m\phi +c0[m,n] \sin m\phi)\right]\,, \label{eq:ur series}
\end{align}
\begin{align}
u_{\theta}=&\frac{(3 + \Gamma)}{8 \pi (2 + \Gamma) \xi}\left( -F_{z} \sin \theta + \cos \theta \left[F_{x} \cos \phi + F_{y} \sin \phi \right]\right)+\frac{1}{8 \pi \xi^2}\left(T_{y} \cos \phi - T_{x} \sin \phi \right)\nonumber \\&+\frac{1}{\sin \theta}\left[\sum_{n=1}^{\infty} \sum_{m=0}^{n}\left(\frac{\xi^n m(2n+1)}{n (n+1)}P_n^m [-a1[m,n] \sin m\phi +a0[m,n] \cos m\phi]\right)\right]\nonumber \\&+
\frac{1}{\sin \theta}\left[\sum_{n=1}^{\infty} \sum_{m=0}^{n}\left( \frac{\xi^{n+1} [\Gamma(n+3)+(n+5)]}{(2n+1)(n+1)[\Gamma\, n +(n-2)]}(n(n-m+1) P_{n+1}^m-(n+1)\right.\right.\nonumber \\& \left.\left.\times(n+m)P_{n-1}^m)[b1[m,n] \cos m\phi +b0[m,n] \sin m\phi]\vphantom{\frac{\xi^{n+1} [\Gamma(n+3)+(n+5)]}{(2n+1)(n+1)[\Gamma\, n +(n-2)]}}\right)\vphantom{\sum_{n=1}^{\infty} \sum_{m=0}^{n}}\right]\nonumber \\
&+\frac{1}{\sin \theta}\left[\sum_{n=1}^{\infty} \sum_{m=0}^{n}\left( \frac{\xi^{n-1}}{(2n+1)n}\left[n(n-m+1) P_{n+1}^m-(n+1)(n+m)P_{n-1}^m\right]\right.\right.\nonumber\\& \left.\left.\times[c1[m,n] \cos m\phi +c0[m,n] \sin m\phi]\vphantom{\frac{\xi^{n-1}}{(2n+1)n}}\right)\vphantom{\sum_{n=1}^{\infty} \sum_{m=0}^{n}}\right],
\end{align}
and
\begin{align}
u_{\phi}= &\frac{(3 + \Gamma)}{8 \pi (2 + \Gamma) \xi} \left(F_{y} \cos \phi - F_{x} \sin \phi \right)+\frac{1}{8 \pi \xi^2} \left(T_{z} \sin \theta - \cos \theta \left[T_{x} \cos \phi + T_{y} \sin \phi \right]\right)\nonumber\\&+\frac{1}{\sin \theta}\left[\sum_{n=0}^{\infty} \sum_{m=0}^{n}\left(\frac{-\xi^n}{(n+1)} \left[(n-m+1) P_{n+1}^m(a1[m,n] \cos m\phi +a0[m,n] \sin m\phi)\right]\right)  \right.\nonumber\\&\left.+\sum_{n=1}^{\infty} \sum_{m=0}^{n}\left(\frac{\xi^n}{n}\left[(n+m)P_{n-1}^m(a1[m,n] \cos m\phi +a0[m,n] \sin m\phi) \right]\right)\vphantom{\sum_{m=0}^{n}\xi^n \left(-\frac{(n-m+1)}{n+1}P_{n+1}^m + \frac{n+m}{n}P_{n-1}^m\right )}\right]\nonumber  \\
&+\frac{1}{\sin \theta}\left[\sum_{n=0}^{\infty} \sum_{m=0}^{n}\left( \frac{\xi^{n+1} m [\Gamma(n+3)+(n+5)]}{(n+1)[\Gamma \, n +(n-2)]} P_{n}^m [-b1[m,n] \sin m\phi \right.\right.\nonumber\\&\left.\left.+b0[m,n] \cos m\phi]\vphantom{\frac{\xi^{n+1} m [\Gamma(n+3)+(n+5)]}{(n+1)[\Gamma \, n +(n-2)]}}\right)\vphantom{\sum_{n=0}^{\infty} \sum_{m=0}^{n}}\right]\nonumber \\
&+\frac{1}{\sin \theta}\left[\sum_{n=1}^{\infty} \sum_{m=0}^{n} \left (\frac{\xi^{n-1} m}{n} P_{n}^m [-c1[m,n] \sin m\phi +c0[m,n] \cos m\phi]\right )\right],\label{eq:uphi series}
\end{align}
respectively. Here, the constants, $a1[m,n]$, $a0[m,n]$, $b1[m,n]$, $b0[m,n]$, $c1[m,n]$, and $c0[m,n]$ are determined at different orders in $\alpha$ until \textit{O}($\alpha^2$) from the corresponding modified stress boundary conditions, as discussed in section \ref{subsubsec:mod stress bc}. The displacement field for the point force decays as $1/\xi$  and for the point torque as $1/\xi^2$ [for details, see Appendix A of \citet{verma2025dynamics}]

\section{Expression of $A_i$,$B_i$,$C_i$,...,$G_i$}\label{appsec:expression of constants from A to G}

\begin{align}
    A_1=&\frac{1}{2}\left[\frac{(1+3(h_2^2-h_1^2+h_1^2 V_{0,z}))}{h_1^2}\cos\theta+3 \sin\theta ( V_{0,x} \cos\phi + V_{0,y} \sin\phi)\right],\\
    A_2=&\frac{5 h_{2} \cos\theta \cos\phi \sin\theta}{h_{1}^{2}},\\
    A_3=&-\frac{\cos\theta( 1 + 4 h_{2}^{2} + 4 h_{1}^{2} (-1 + V_{0,z}) + 7 \cos2\theta - 14 \cos2\phi \sin^{2}\theta )}{8 h_{1}^{2}}\nonumber\\&-\frac{1}{2} \sin\theta ( V_{0,x} \cos\phi + V_{0,y} \sin\phi ),\\
    A_4=&-\frac{3 A_{2}}{5}, A_5=\frac{1}{8 h_{1}^{2}}\left[-1 + 5 \cos2\theta - 10 \cos2\phi \sin^{2}\theta\right]\cos\theta,
\end{align}
\begin{align}
    B_1=&\frac{-1}{4 h_1^2}\left[(1 + 3 h_{2}^{2} + 3 h_{1}^{2} (-1 + V_{0,z})) \sin\theta\right] + \frac{3}{4}\cos\theta\left[V_{0,x} \cos\phi + V_{0,y} \sin\phi\right],\\
    B_2=&-\frac{h_{2} \cos\phi}{h_{1}^{2}},\\ B_3=&\frac{-1}{32 h_1^2}\left[(9 + 8 h_{2}^{2} + 8 h_{1}^{2} (-1 + V_{0,z}) + 14 \cos2\theta \cos^{2}\phi + 13 \cos2\phi ) \sin\theta\right]\nonumber\\&+\frac{1}{4}\cos\theta\left[V_{0,x} \cos\phi + V_{0,y} \sin\phi\right],\\
    B_4=&\frac{h_{2} \cos2\theta \cos\phi}{h_{1}^{2}}, B_5=\frac{1}{64 h_1^2}\left[(3 - 5 \cos2\phi) \sin\theta + 30 \cos^{2}\phi \sin3\theta \right],
\end{align}
\begin{align}
    C_1=&\frac{3}{4}\left[V_{0,y} \cos\phi - V_{0,x} \sin\phi\right],
    C_2=\frac{h_{2} \cos\theta \sin\phi}{h_{1}^{2}} ,\\
    C_3=&\frac{1}{4}\left[- V_{0,x} \sin\phi + \cos\phi \left( V_{0,y} + \frac{5 \cos\theta \sin\theta \sin\phi}{h_{1}^{2}} \right)\right] ,\\
    C_4=& -C_2,
    C_5=-\frac{5 \cos\theta \cos\phi \sin\theta \sin\phi}{4 h_{1}^{2}},
\end{align}

\begin{align}
    D_1=& A_1 ,  D_2=2 A_2  , D_3= \frac{1}{16 h_{1}^{2}}\left[7 \cos\theta (1 - 5 \cos2\theta + 10 \cos2\phi \sin^{2}\theta)\right] ,
\end{align}

\begin{align}
    E_1=&-\frac{P_{0}}{2 + 3 \Gamma} - \frac{5 h_{2} \cos\theta \cos\phi \sin\theta}{h_{1}^{2}},\\
    E_2=&\frac{1}{64 h_{1}^{2} (2 + \Gamma) (2 + 3 \Gamma)}\left[(-3 (-2 + 3 \Gamma) (22 + 13 \Gamma) - 96 h_{2}^{2} (-1 + \Gamma^{2}) - 96 h_{1}^{2} (-1 + V_{0,z}) \right.\nonumber\\&\left. \times(-1 + \Gamma^{2}) + 35 (2 + \Gamma) (2 + 3 \Gamma) \cos2\theta - 70 (2 + \Gamma) (2 + 3 \Gamma) \cos2\phi \sin^{2}\theta)\cos\theta \right.\nonumber\\&\left.- 96 h_{1}^{2} (-1 + \Gamma^{2}) \sin\theta (V_{0,x} \cos\phi + V_{0,y} \sin\phi) \right],\\
    E_3=&A_1,
\end{align}

\begin{align}
    F_1=&-\frac{5}{2}B_4,\\
    F_2=&\frac{1}{128 h_{1}^{2} (2 + \Gamma) (2 + 3 \Gamma)}\left[(932 + 192 h_{2}^{2} (1 + \Gamma) (3 + 2 \Gamma) + 192 h_{1}^{2} (-1 + V_{0,z}) (1 + \Gamma)\right.\nonumber\\&\left.\times (3 + 2 \Gamma) + \Gamma (1160 + 363 \Gamma) + 225 (2 + \Gamma) (2 + 3 \Gamma) \cos2\phi)\sin\theta-  70 (2 + \Gamma) \right.\nonumber\\&\left.\times(2 + 3 \Gamma) \cos^{2}\phi \sin3\theta - 192 h_{1}^{2} (1 + \Gamma) (3 + 2 \Gamma) \cos\theta (V_{0,x} \cos\phi + V_{0,y} \sin\phi) \right],\\
    F_3=& \frac{3+\Gamma}{2+\Gamma}\,B_1  , F_4=B_2,
\end{align}

\begin{align}
    G_1=&\frac{5 h_{2} \cos\theta \sin\phi}{2 h_{1}^{2}} ,\\
    G_2=&\frac{3}{8}\left[-\frac{5 \cos\theta \cos\phi \sin\theta \sin\phi}{h_{1}^{2}}+\frac{4 (1 + \Gamma) (3 + 2 \Gamma) (-V_{0,y} \cos\phi + V_{0,x} \sin\phi)}{(2 + \Gamma) (2 + 3 \Gamma)}\right] ,\\
    G_3=&\frac{3 (3 + \Gamma) (V_{0,y} \cos\phi - V_{0,x} \sin\phi)}{4 (2 + \Gamma)}, G_4=\frac{2}{5}G_1,
\end{align}

\section{Expression of $S_i$, where $i=1,2,...,6$ and of $(S_3)_{V_{0,z}}$}\label{appsec:S1 to S6 expression}

\begin{align}
    S_1=&\frac{V_{0,x}}{13023360 h_1^4}\left[5(226616 + \frac{1281987}{2 + \Gamma} - \frac{4069800}{(2 + 3 \Gamma)^{2}} - \frac{1316700}{14 + 17 \Gamma} - \frac{21718809}{14 + 19 \Gamma}) \right.\nonumber\\&\left.+ \frac{15504}{(2 + \Gamma) (2 + 3 \Gamma)^{2} (14 + 19 \Gamma) (26 + 33 \Gamma)} ( 1505 (2 + \Gamma) (2 + 3 \Gamma) (14 + 19 \Gamma) (26 + 33 \Gamma)\right.\nonumber\\&\left.+ 945 h_{2}^{4} (26 + 33 \Gamma) (-140 + \Gamma (-360 + \Gamma (-227 + \Gamma (6 + \Gamma)))) + 945 h_{1}^{4} (V_{0,x}^{2} + V_{0,y}^{2} \right.\nonumber\\&\left.+ (-1 + V_{0,z})^{2}) (26 + 33 \Gamma) (-140 + \Gamma (-360 + \Gamma (-227 + \Gamma (6 + \Gamma))))+ 315 h_{1}^{2} (-1 \right.\nonumber\\&\left.+ V_{0,z}) (26 + 33 \Gamma) (-364 + \Gamma (-444 + \Gamma (475 + \Gamma (653 + 130 \Gamma))) + 6 h_{2}^{2} (-140 \right.\nonumber\\&\left.+ \Gamma (-360 + \Gamma (-227 + \Gamma (6 + \Gamma))))) + h_{2}^{2} (7164696 + 3 \Gamma (14136700 + \Gamma (30534634 \right.\nonumber\\&\left.+ 3 \Gamma (9874779 + \Gamma (4103053 + 503571 \Gamma)))))        )   \vphantom{\frac{1316700}{14 + 17 \Gamma}}  \right]
\end{align}

\begin{align}
    S_2=&\frac{V_{0,y}}{(20160 h_{1}^{4} (2 + \Gamma) (2 + 3 \Gamma)^{2} (14 + 17 \Gamma) (14 + 19 \Gamma) (26 + 33 \Gamma))} \left[\vphantom{\frac{V_{0,y}}{(20160 h_{1}^{4} (2 + \Gamma) (2 + 3 \Gamma)^{2} (14 + 17 \Gamma) (14 + 19 \Gamma) (26 + 33 \Gamma))}}   22680 h_{2}^{4} (14 + 17 \Gamma) \right.\nonumber\\&\left.\times(26 + 33 \Gamma) (-140 + \Gamma (-360 + \Gamma (-227 + \Gamma (6 + \Gamma)))) + 22680 h_{1}^{4} (V_{0,x}^{2} + V_{0,y}^{2} \right.\nonumber\\&\left.+ (-1 + V_{0,z})^{2}) (14 + 17 \Gamma) (26 + 33 \Gamma) (-140 + \Gamma (-360 + \Gamma (-227 + \Gamma (6 + \Gamma))))
\right.\nonumber\\&\left.- 5 (26 + 33 \Gamma) (10834096 + \Gamma (50991584 + \Gamma (93788360 + \Gamma (84095872 + 36563403 \Gamma \right.\nonumber\\&\left.+ 6064830 \Gamma^{2})))) +  3780 h_{1}^{2} (-1 + V_{0,z}) (14 + 17 \Gamma) (26 + 33 \Gamma) (-1288 - \Gamma (3604 \right.\nonumber\\&\left.+ \Gamma (3794 + \Gamma (2083 + 481 \Gamma))) + 12 h_{2}^{2} (-140 + \Gamma (-360 + \Gamma (-227 + \Gamma (6 + \Gamma))))) \right.\nonumber\\&\left.+ 36 h_{2}^{2} (14 + 17 \Gamma) (8263472 + \Gamma (46286000 + \Gamma (96483224 + 3 \Gamma (30835904 \right.\nonumber\\&\left.+ \Gamma (13395713 + 2118201 \Gamma)))))    \vphantom{\frac{V_{0,y}}{(20160 h_{1}^{4} (2 + \Gamma) (2 + 3 \Gamma)^{2} (14 + 17 \Gamma) (14 + 19 \Gamma) (26 + 33 \Gamma))}}    \right]
\end{align}

\begin{align}
    S_3=&\frac{1}{(221760 h_{1}^{6} (2 + \Gamma)^{2} (2 + 3 \Gamma)^{2} (14 + 17 \Gamma) (14 + 19 \Gamma) (26 + 33 \Gamma))} \left[ \vphantom{\frac{1}{(221760 h_{1}^{6} (2 + \Gamma)^{2} (2 + 3 \Gamma)^{2} (14 + 17 \Gamma) (14 + 19 \Gamma) (26 + 33 \Gamma))}} 249480 h_{1}^{6} (V_{0,x}^{2} \right.\nonumber\\&\left.+ V_{0,y}^{2} + (-1 + V_{0,z})^{2}) (-1 + V_{0,z}) (2 + \Gamma) (14 + 17 \Gamma) (26 + 33 \Gamma) (-140 + \Gamma (-360 \right.\nonumber\\&\left.+ \Gamma (-227 + \Gamma (6 + \Gamma)))) + 6930 h_{1}^{4} (14 + 17 \Gamma) (26 + 33 \Gamma) ( 36 h_{2}^{2} (V_{0,x}^{2} + V_{0,y}^{2} \right.\nonumber\\&\left.+ 3 (-1 + V_{0,z})^{2}) (2 + \Gamma) (-140 + \Gamma (-360 + \Gamma (-227 + \Gamma (6 + \Gamma))))
- 24 V_{0,x}^{2} (686 \right.\nonumber\\&\left.+ \Gamma (2819 + 10 \Gamma (450 + \Gamma (361 + \Gamma (151 + 24 \Gamma))))) + 3 (2 + \Gamma) ((-1 + V_{0,z})^{2} (-2520 \right.\nonumber\\&\left.+ \Gamma (-4244 + \Gamma (234 + \Gamma (1893 + 47 \Gamma)))) + V_{0,y}^{2} (56 + \Gamma (2964 + \Gamma (7822 + \Gamma (6059 \right.\nonumber\\&\left.+ 1009 \Gamma)))))   ) +  99 h_{1}^{2} (-1 + V_{0,z}) (2 + \Gamma) ( 7560 h_{2}^{4} (14 + 17 \Gamma) (26 + 33 \Gamma) (-140 \right.\nonumber\\&\left.+ \Gamma (-360 + \Gamma (-227 + \Gamma (6 + \Gamma)))) - 5 (26 + 33 \Gamma) (3047408 + \Gamma (14243936 \right.\nonumber\\&\left.+ \Gamma (27918120 + \Gamma (29116864 + \Gamma (15872575 + 3404742 \Gamma))))) - 12 h_{2}^{2} (14 + 17 \Gamma) \right.\nonumber\\&\left.\times(4590992 + \Gamma (21084800 + \Gamma (41579904 + \Gamma (45565132 + \Gamma (27288199 \right.\nonumber\\&\left.+ 6653973 \Gamma)))))  )  + (2+\Gamma) (249480 h_{2}^{6} (14 + 17 \Gamma) (26 + 33 \Gamma) (-140 + \Gamma (-360 \right.\nonumber\\&\left.+ \Gamma (-227 + \Gamma (6 + \Gamma)))) - 5 (2 + \Gamma) (26 + 33 \Gamma) (142225832 + \Gamma (608291292 \right.\nonumber\\&\left.+ \Gamma (980502694 + 3 \Gamma (238380523 + 67513296 \Gamma)))) - 594 h_{2}^{4} (14 + 17 \Gamma) (6888784 \right.\nonumber\\&\left.+ \Gamma (35396960 + \Gamma (78470928 + \Gamma (93123164 + \Gamma (56805583 + 13362231 \Gamma))))) \right.\nonumber\\&\left.+ 99 h_{2}^{2} (1679266848 + \Gamma (11905184144 + \Gamma (33065352016 + \Gamma (45873575416 \right.\nonumber\\&\left.+ \Gamma (33189937250 + 9 \Gamma (1298256065 + 171411874 \Gamma))))))  ) \vphantom{\frac{1}{(221760 h_{1}^{6} (2 + \Gamma)^{2} (2 + 3 \Gamma)^{2} (14 + 17 \Gamma) (14 + 19 \Gamma) (26 + 33 \Gamma))}} \right]
\end{align}

\begin{align}
    (S_3)_{V_{0,z}}=&S_3-\frac{1}{32 h_1^2 (2+\Gamma)^2 (2+3\Gamma)^2 (14+19\Gamma)}\left[\vphantom{\frac{1}{32 h_1^2 (2+\Gamma)^2 (2+3\Gamma)^2 (14+19\Gamma)}}3(V_{0,y}^2(2+\Gamma)(56+\Gamma(2964\right.\nonumber\\&\left.+\Gamma(7822+\Gamma(6059+1009\Gamma)))+12h_2^2(-140+\Gamma(-360+\Gamma(-227\right.\nonumber\\&\left.+\Gamma(6+\Gamma))))+12 h_1^2(-1+V_{0,z})(-140+\Gamma(-360+\Gamma(-227+\Gamma(6+\Gamma))))) \right.\nonumber\\&\left.+ 4 V_{0,x}^2 (3 h_2^2 (2+\Gamma)(-140+\Gamma(-360+\Gamma(-227+\Gamma(6+\Gamma)))) + 3 (h_1^2) (-1\right.\nonumber\\&\left.+V_{0,z}) (2+\Gamma) (-140+\Gamma(-360+\Gamma(-227+\Gamma(6+\Gamma)))) - 2(686+\Gamma(2819\right.\nonumber\\&\left.+10\Gamma(450+\Gamma(361+\Gamma(151+24\Gamma))))) )  )\vphantom{\frac{1}{32 h_1^2 (2+\Gamma)^2 (2+3\Gamma)^2 (14+19\Gamma)}}\right] \label{appeq:S_3_V0z}
\end{align}

\begin{align}
    S_4=-\frac{9 h_{2} V_{0,x} V_{0,y} (5432 + \Gamma (20072 + \Gamma (24820 + \Gamma (11081 + 1055 \Gamma))))}{32 h_{1}^{2} (2 + \Gamma) (2 + 3 \Gamma)^{2} (14 + 19 \Gamma)}
\end{align}

\begin{align}
    S_5=&\frac{1}{(6720 h_{1}^{6} (2 + \Gamma) (2 + 3 \Gamma)^{2} (14 + 19 \Gamma) (26 + 33 \Gamma))} \left[\vphantom{\frac{1}{(6720 h_{1}^{6} (2 + \Gamma) (2 + 3 \Gamma)^{2} (14 + 19 \Gamma) (26 + 33 \Gamma))}} 1890 h_{2}^{5} (26 + 33 \Gamma) (2128 \right.\nonumber\\&\left.+ \Gamma (5792 + \Gamma (4902 + 7 \Gamma (257 + 97 \Gamma)))) + 12 h_2^3 (-66233216 - \Gamma (361607720 \right.\nonumber\\&\left.+ \Gamma (734452472 + 9 \Gamma (75764974 + 31395528 \Gamma + 4621461 \Gamma^{2}))) + 315 h_{1}^{2} (-1 + V_{0,z})\right.\nonumber\\&\left.\times (26 + 33 \Gamma) (2128 + \Gamma (5792 + \Gamma (4902 + 7 \Gamma (257 + 97 \Gamma))))  ) + 5 h_2 (16 (1733186 \right.\nonumber\\&\left.+ 6305375 \Gamma) + \Gamma^{2} (141509912 + 3 \Gamma (36172648 + 3 \Gamma (5977492 + 1367385 \Gamma))) \right.\nonumber\\&\left.+ 378 h_{1}^{4} (26 + 33 \Gamma) ( -12 V_{0,y}^{2} (-140 + \Gamma (-360 + \Gamma (-227 + \Gamma (6 + \Gamma)))) + (-1 \right.\nonumber\\&\left.+ V_{0,z})^{2} (2128 + \Gamma (5792 + \Gamma (4902 + 7 \Gamma (257 + 97 \Gamma)))) + V_{0,x}^{2} (7112 + \Gamma (24392 \right.\nonumber\\&\left.+ \Gamma (27544 + \Gamma (11009 + 1043 \Gamma))))  ) + 36 h_{1}^{2} (-1 + V_{0,z}) (1821568 + \Gamma (8862120 \right.\nonumber\\&\left.+ \Gamma (17853816 + \Gamma (18790718 + 3 \Gamma (3394712 + 720639 \Gamma)))))  ) \vphantom{\frac{1}{(6720 h_{1}^{6} (2 + \Gamma) (2 + 3 \Gamma)^{2} (14 + 19 \Gamma) (26 + 33 \Gamma))}} \right]
\end{align}

\begin{align}
    S_6=&\frac{1}{(224 h_{1}^{4} (2 + \Gamma) (2 + 3 \Gamma)^{2} (14 + 19 \Gamma) (26 + 33 \Gamma))} \left[ \vphantom{\frac{1}{(224 h_{1}^{4} (2 + \Gamma) (2 + 3 \Gamma)^{2} (14 + 19 \Gamma) (26 + 33 \Gamma))}} 3 h_2 V_{0,y} (315840 - 21 h_{2}^{2} (26 + 33 \Gamma)\right.\nonumber\\&\left.\times (448 + \Gamma (1472 + \Gamma (2178 + \Gamma (1871 + 691 \Gamma)))) - 21 h_{1}^{2} (-1 + V_{0,z}) (26 + 33 \Gamma) (448 \right.\nonumber\\&\left.+ \Gamma (1472 + \Gamma (2178 + \Gamma (1871 + 691 \Gamma)))) + \Gamma (2359184 + \Gamma (5657948 + \Gamma (5699684 \right.\nonumber\\&\left.+ 9 \Gamma (262903 + 34183 \Gamma)))) )  \vphantom{\frac{1}{(224 h_{1}^{4} (2 + \Gamma) (2 + 3 \Gamma)^{2} (14 + 19 \Gamma) (26 + 33 \Gamma))}} \right]
\end{align}
\bibliographystyle{jfm}
\bibliography{jfm}
\end{document}